\documentclass[%
 reprint,
 superscriptaddress,
 amsmath,amssymb,
 aps,
pra]{revtex4-2}

\usepackage{graphicx}
\usepackage{dcolumn}
\usepackage{xcolor}
\usepackage{bm}
\usepackage{hyperref}
\usepackage{lipsum}
\usepackage{braket}
\usepackage{amsmath}
\usepackage{physics}
\usepackage[english]{babel}
\usepackage{natbib}

\newcommand{\ii}{\mathrm{i}}

\newcommand{\ee}{\mathrm{e}}

\renewcommand{\dd}{\mathrm{d}}

\newcommand{\ve}[1]{\boldsymbol{#1}}

\begin{document}

\newcommand{\red}[1]{\textcolor{red}{#1}}

\preprint{APS/123-QED}

\title{Ultra-slow orbital and spin dynamics in an electrically tunable quantum dot molecule}

\author{Christopher Thalacker}
\email[Corresponding author: ]{christopher.thalacker@tum.de}
\author{Michelle Lienhart}%
\author{Markus St\"ocker}
\author{Nadeem Akhlaq}
\author{Irina Ivanova}

\affiliation{
 Walter-Schottky-Institute, Technische Universit\"at M\"unchen, Am Coulombwall 4, 85748 Garching, Germany
}

\author{Nikolai Bart}
\author{Arne Ludwig}
\affiliation{Faculty of Physics and Astronomy, Ruhr-Universit\"at Bochum, Universit\"atsstr. 150, 44801 Bochum, Germany
}

\author{Johannes Schall}
\author{Stephan Reitzenstein}
\affiliation{%
 Institute for Physics and Astronomy, Technische Universit\"at Berlin, Hardenbergstr. 36, 10623 Berlin, Germany
}%

\author{Dirk Reuter}
\affiliation{%
  Universit\"at Paderborn, Warburger Str. 100, 33098 Paderborn, Germany
}%

\author{Steffen Wilksen}
\author{Christopher Gies}
\affiliation{%
  Carl von Ossietzky Universit\"at Oldenburg, Fakult\"at V, Institut f\"ur Physik, 26129 Oldenburg, Germany
}%

\author{Krzysztof Gawarecki}
\author{Pawe\l{} Machnikowski}
\affiliation{%
Institute of Theoretical Physics, Wroc\l{}aw University of Science and Technology, Wroc\l{}aw 50-370, Poland
}%

\author{Kai M\"uller}
\affiliation{%
  TUM School of Computation, Information and Technology, TU M\"unchen, Hans-Piloty-Str. 1, 85748 Garching, Germany
}%

\author{Jonathan Finley}
\affiliation{
 Walter-Schottky-Institute, Technische Universit\"at M\"unchen, Am Coulombwall 4, 85748 Garching, Germany
}

\date{\today}

\begin{abstract}
Tunnel-coupled optically active quantum dot molecules (QDMs), have the potential to operate as spin-photon-interfaces with coupled spins that interact with two different photon frequencies at the same time. A prerequisite is to deterministically prepare two (electron or hole) spins in the QDM and be able to electrically tune the orbital state couplings. Here, we demonstrate the sequential optical charging of a single QDM with two electron spins while simultaneously maintaining the ability to widely tune orbital couplings using static electric fields and optically drive the system for quantum light generation. We optically prepare one- and two-spin states, initialize via optical pumping and explore orbital and spin relaxation dynamics for one and two-spin states as a function of the energy detuning and hybridization of orbital states. For two-spin states, remarkably long S-T relaxation times are observed extending beyond $\sim 100\mu s$ with strong dependence on the relative energy of ground and excited two-spin states.  Qualitative agreement is observed with $\mathbf{k \cdot p}$ calculations of phonon-mediated spin-relaxation. Our results provide new quantitative understanding of the dynamics of one and two-spin states and confirm their suitability of QDMs for creating multidimensional photonic cluster states by exploiting tunable spin-spin exchange couplings at zero magnetic fields combined with optical driving.
\end{abstract}

\maketitle

\section{\label{sec:introduction}Introduction}
Spin-photon interfaces can exploit light-matter interactions to generate non-classical states of light and mediate entanglement between non-interacting photonic qubits encoded in polarization, time or path degrees of freedom. In this context, optically active semiconductor quantum dots represent the gold-standard for non-classical light sources. They are excellent on-demand single photon emitters\,\cite{awschalom_quantum_2018, hanschke_quantum_2018, thomas_bright_2021}, can produce high-fidelity entangled photon pairs\,\cite{akopian_entangled_2006} and serve as reliable spin-photon interfaces\,\cite{press_complete_2008, gao_observation_2012, stockill_phase-tuned_2017}. Recently, single dots have been used to deterministically generate extended, one-dimensional (1D) linear chains of entangled photons, so-called photonic cluster states\,\cite{lindner_proposal_2009, schwartz_deterministic_2015, cogan_deterministic_2023, coste_high-rate_2023}.
However, multi-photon states with 2D entanglement structures are required for measurement-based quantum information processing\,\cite{raussendorf_measurement-based_2003}. Such states can be generated from 1D cluster states using probabilistic fusion gates\,\cite{browne_resource-efficient_2005}, a process that does not scale as well with cluster state size. An alternative approach for deterministically generating cluster states with 2D entanglement structures is to use a pair of vertically stacked and coupled quantum dots, a quantum dot molecule (QDM) containing two interacting spins\,\cite{economou_optically_2010, gimeno-segovia_deterministic_2019}. In these structures, the dots are separated by a thin tunnel barrier that couples the orbital states of trapped electrons or holes\,\cite{krenner_direct_2005, doty_optical_2008}. When populated by two spins, the hybridized orbital states promise longer spin coherence times due to the formation of noise-protected singlet-triplet (S-T) qubits in the two-spin subspace\,\cite{bluhm_dephasing_2011, weiss_coherent_2012, tran_enhanced_2022, delley_deterministic_2017}.  This facilitates fast entangling operations on the two spins that can then be mapped into the photonic domain by optical driving at well defined times\,\cite{economou_optically_2008, kim_ultrafast_2011, greilich_optical_2011}. Using QDMs for such applications requires the ability to (i) deterministically load the structure with two spins (either electron or hole) and operate in a stable charge occupancy regime and (ii) control the energetic detuning between orbitals in both dots. These requirements can each be individually fulfilled by embedding the QDM into the intrinsic region of a p-i-n diode and charging it using Coulomb blockade from a tunnel coupled reservoir. However since a single control parameter -- the axial electric field -- is used to tune both orbital coupling and charge status, it has proven difficult to realize two-spin systems in widely tunable, optically active QDMs.\\

An alternative approach involves using charge storage devices\,\cite{bopp_quantum_2022} in which spins are \textit{optically} prepared by resonant optical driving\,\cite{bopp_quantum_2022,bopp_electrical_2023,bopp_coherent_2023}. There, we demonstrate selective optical charging of a single QDM with one or two \textit{electron spins}. After preparation, we show that one or two spins can be readily tuned into a regime where the orbital wavefunctions in the two dots hybridize,  forming states with mixed spin and orbital character. Moreover, we show that spin states in the QDM can be optically pumped over nanosecond timescales and we measure the orbital and spin relaxation processes as the orbital coupling is electrically varied. Our results show that hybridized states can be repeatedly optically addressed before any charge loss occurs, with charges remaining in the QDM over timescales $\gg50\,\mathrm{\mu s}$ at 1.7\,K. For the two-spin system, we show that it can be initialized into a well-defined singlet ground state that maintains its character even after electrical switching. We use our charge storage device to elucidate the relaxation dynamics of single and two-electron spin states under optical pumping.  The S-T relaxation is shown to be very slow for energy splittings $\Delta E_{S-T} < 5\,\mathrm{meV}$, with relaxation rates as low as $0.01\,\mathrm{\mu s}^{-1}$.   Remarkably, upon increasing  $\Delta E_{S-T}$ beyond 5\,meV, the relaxation rate suddenly increases by more than two-orders of magnitude due to the onset of sequential multi-phonon spin relaxation via excited state triplets.

\section{\label{sec:plv}Device structure, optical charging and single QDM emission}
\begin{figure*}[t]
    \includegraphics[width=0.9\textwidth]{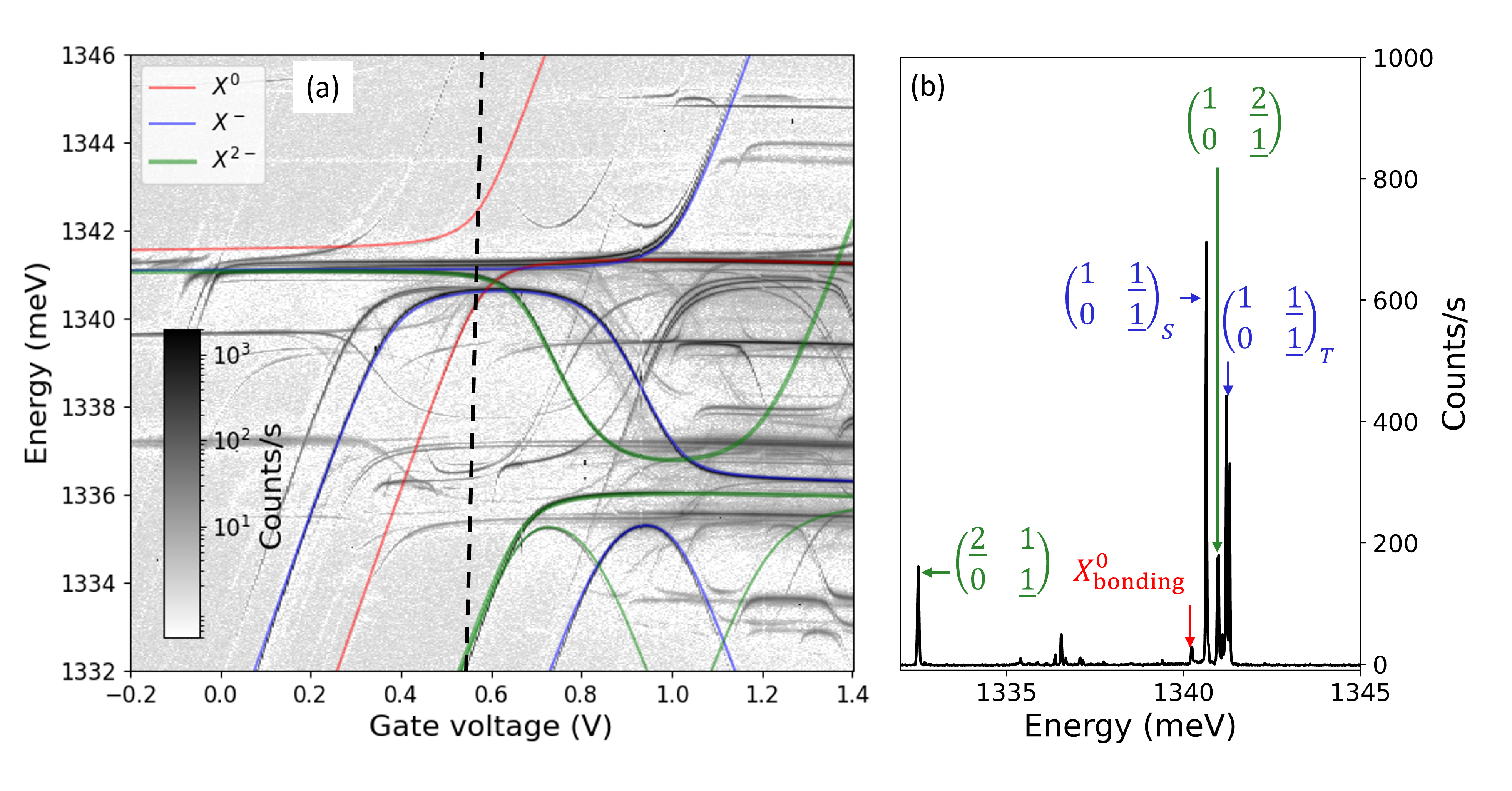}
    \caption{\label{fig:plv}Voltage-dependent photoluminescence. (a) Voltage-dependent photoluminescence spectrum of a single quantum dot molecule. Different charged excitonic complexes are identified and overlaid with the spectrum in different colors. (b) PL-spectrum of the QDM at $V_G=0.56\,\mathrm{V}$ (marked by the dashed line in (a)). Main transitions are labeled according to the convention explained in the main text.}
\end{figure*}

Our devices consist of a single self-assembled QDM formed by a pair of vertically stacked InGaAs dots with a $\sim$ 10\,nm dot-to-dot separation. The QDM is embedded into the intrinsic region of a p-i-n diode to apply static electric fields along the growth axis. The layer structure is designed for electron storage:  The lower dot is flatter (larger confinement energy) than the upper quantum dot, and a 50\,nm thick AlGaAs barrier below the lower dot prevents the electrons from escaping (see also supplementary information\,\ref{sec:sample_design}). A circular Bragg grating (CBG) is deterministically fabricated directly above the QDM using in-situ electron-beam lithography\,\cite{rodt_high-performance_2021}. In this process, low-temperature cathodoluminescence is first employed to locate the molecule, after which electron-beam lithography defines and precisely aligns the CBG to the selected QDM\,\cite{schall_bright_2021}. In combination with a backside distributed Bragg reflector, this approach increases the efficiency of photon collection over a $\sim$ 10\,nm bandwidth\, sufficient to simultaneously optically address transitions in the upper and lower dot\,\cite{schall_bright_2021}. Initially, we spectroscopically probe the orbital states of neutral and charged excitons in the QDM by performing voltage-dependent photoluminescence (PL-V) with a continuous-wave (cw) laser at 850\,nm exciting quasi-free charge carriers in the wetting layer. For these PL-V measurements we prevent the accumulation of electrons in the QDM over time by applying a negative gate voltage $V_G$ periodically to remove accumulated photogenerated electrons. This discharging phase is followed by a readout phase in which the QDM is excited. Fig.\,\ref{fig:plv}(a) shows a typical example of such PL-V measurementS. The data shows strongly and weakly shifting lines that indicate the presence of different indirect and direct excitonic complexes, along with avoided crossings as reported previously for QDMs\,\cite{bayer_coupling_2001,krenner_direct_2005,greilich_optical_2011}. Different charged complexes in the QDM can be readily identified by fitting the data using a few-body Hamiltonian (see supplementary information\,\ref{sec:charge_states}). The model provides a consistent fit to the data, revealing neutral exciton states $X^0$ (red), singly charged excitons $X^-$ (blue) and doubly charged excitons $X^{2-}$ (green) in the QDM. We further characterize the transitions via the orbital configuration of electrons and the holes, adopting the state labeling convention used in Ref.\,\cite{doty_optical_2008}. Here, the orbital configuration is denoted by a $2 \times 2$ matrix of the form
\begin{equation}
    \begin{pmatrix}
        e_l & e_u \\
        h_l & h_u
    \end{pmatrix}
\end{equation}
where $e_l$ ($e_u$) is the number of electrons in the lower (upper) dots and $h_l$ ($h_u$) is the number of holes in the lower (upper) dots, respectively. When describing transitions, the underscores indicate which electron and hole recombine to generate the resulting photon. Fig.\,\ref{fig:plv}(b) shows a selected spectrum recorded from the QDM at a gate voltage of $V_G=0.56\,\mathrm{V}$ and some of the most important transitions are labeled.\\

For states not involving a hole (i.e. the bare one and two electron states), we also use a shorthand notation in the form of a row vector ($e_l, e_u$) where only the number of electrons in each dot are represented. Where necessary, the spin state of the two-electron-system is denoted as a suffix to the orbital configuration of the form $(e_l, e_u)_s$ where $s \in \{S,T_0,T_+, T_-\}$  representing the symmetry of the two-spin states. In the results presented below, we discriminate between, different voltage regimes. At low voltages, $V_G < 0.7$\,V, the two-electron configuration ($e_L, e_U$)=(2,0) is energetically split off from the (1,1)-configuration by several millielectronvolts. We denote this to be the (2,0)-regime. Between $V_G=0.8\,\mathrm{V}$ and 1.2\,V the lowest two-electron states are (1,1) states hybridized over both dots, forming singlet and triplet spin states (a more detailed description will be given in section\,\ref{sec:singlet-triplet}). We label this as the (1,1)-regime where single dot orbitals are hybridized. The regime where the (0,2)-configuration forms the ground state ($V_G > 1.4$\,V) is inaccessible in our device, since current begins to flow in the diode.

\section{\label{sec:charging}Deterministic two-electron charging}
\begin{figure*}[t]
    \includegraphics[width=0.9\linewidth]{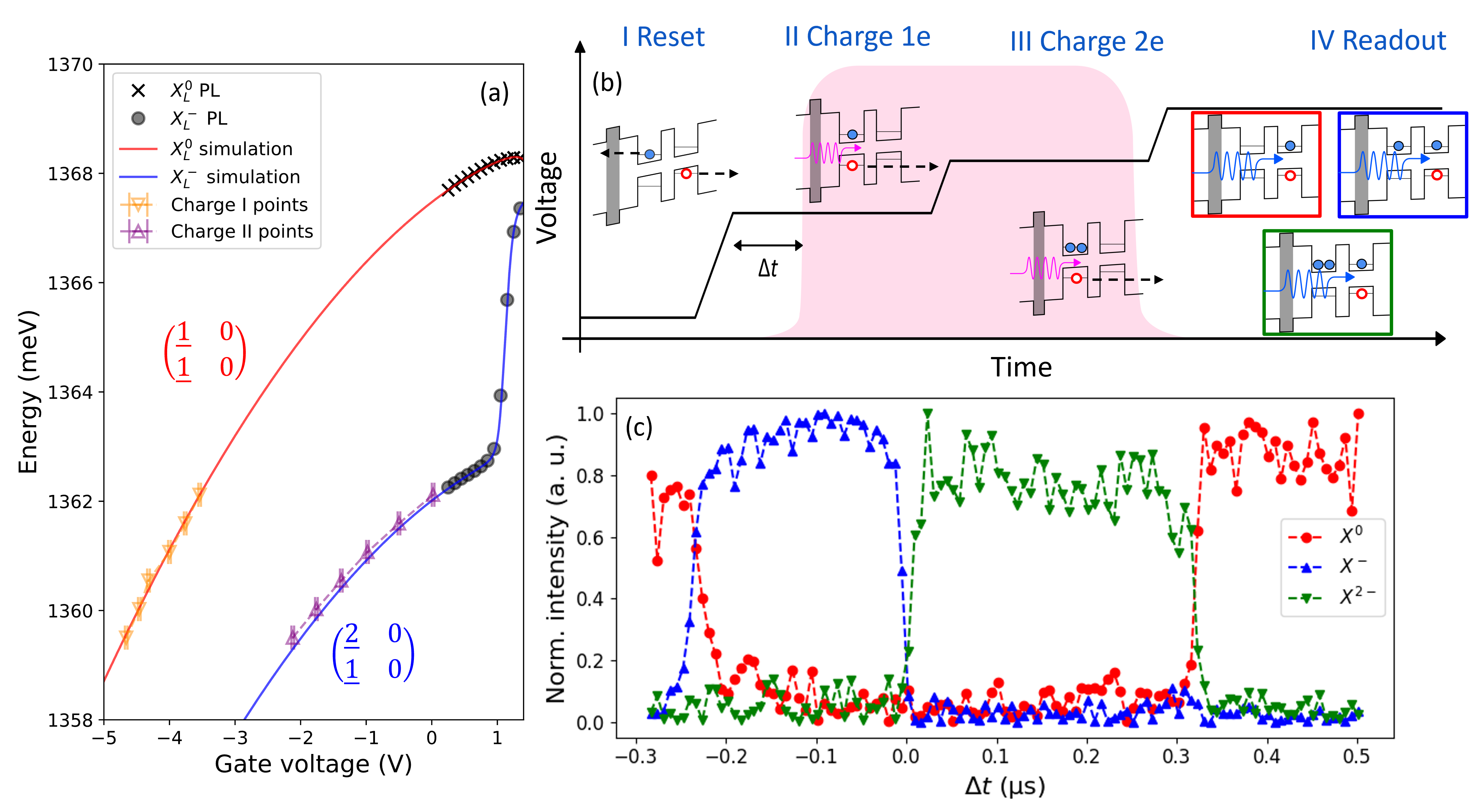}
    \caption{\label{fig:charging_protocol}Deterministic optical electron charging. (a) Voltage and energy point where 1e and 2e charging is possible. In the PL regime ($V_G > 0$\,V) points where luminescence from the lower dot is directly observed, are plotted for the $X^0$ and the $X^-$, along with a few-particle Hamiltonian simulation for these charged complexes. In the regime below 0\,V, points where 1e cand 2e charging occurs (experimental data) are also marked. (b) Schematic of the sequential charging experiment. The charging laser pulse (indicated in magenta) is temporally scanned over the different voltage steps by an offset $\Delta t$ that is measured relative to the start of the first charging plateau. The steps of the charging protocol are discussed in the main text. (c) RF signal of $X^0$, $X^-$ and $X^{2-}$ vs. the time offset of the charge laser pulse relative to the first charge voltage plateau.}
\end{figure*}
The conventional method used to control the charge status of optically active quantum dots is to exploit Coulomb blockade between charges populating the dot and proximal Fermi reservoir\,\cite{warburton_single_2013}. Such stable two-spin regimes are typically achieved over a small region of the gate voltage space, defined by the orbital quantization energies (relative sizes) of the upper and lower dots, their relative In-Ga composition as well as the distance of the QDM to the contact. This complicates free control of the detuning between the orbital states, without precise and independent control over geometry and composition. Here, we adopt an alternative approach that facilitates free tuning of the orbital energy separation in the two dots, whilst retaining two electron spins within the system.\\

Hereby, we utilize an all-optical charging protocol that we previously demonstrated for single dots\,\cite{bechtold_quantum_2016} and holes\,\cite{bopp_quantum_2022} in QDMs. In this charge control scheme, the controlled number of electrons in the molecule is defined by the four stage protocol presented in Fig.\,\ref{fig:charging_protocol}(b). In step I (Reset), a large electric field ($> 140\,\mathrm{kV/cm^{-1}}$) is applied that removes all residual charges by tunneling. Subsequently, in step II, the electric field is set such that the hole tunnels out faster than the exciton radiative lifetime and a resonant laser excites a neutral exciton in the lower dot of the molecule. This way, the QDM is deterministically charged by a single electron with near-unity efficiency. In step III, the same process can be repeated with the resonant laser tuned to the singly charged exciton $X^{-}$. Alternatively, the resonant charging laser can be fixed and the electric field tuned to induce DC Stark shifts to sequentially tune the QDM into resonance with $X^-$. Finally, in step IV (Readout), the electric field is reduced and the induced charge and spin states can be probed using resonance fluorescence (RF) as a function of the orbital coupling defined by $V_G$. Following the method presented in Ref.\,\cite{bopp_quantum_2022}, we first determined the optimal parameters for sequential optical charging by mapping out the parameter space of charging voltage and charge laser energy. Full details are presented in section~\ref{sec:charge_plateaus} of the supplemental. Fig.\,\ref{fig:charging_protocol}(a) shows different pairs of charging energy and voltage for one-electron charging (orange) and two-electron charging (purple). By tracing the voltage-dependence and comparing with the photoluminescence observed from charge states, where the hole is located in the lower dot (shown in Fig.\,\ref{fig:charging_protocol}(a)), the excitonic states via which charging occurs can be identified as $\begin{pmatrix}
    1 & 0\\ 1 & 0
\end{pmatrix}$ for one-electron spin charging and $\begin{pmatrix}
    2 & 0\\ 1 & 0
\end{pmatrix}$ for generation of the second spin. These attributions are confirmed by fitting the Hamiltonian to trace out the field dependent excitonic states (see section \,\ref{sec:charge_states} in the supplemental). As shown in Fig.\,\ref{fig:charging_protocol}(a), very good agreement is observed with calculations of the voltage-dependent eigenenergies for the lower dot of the neutral exciton and singly charged exciton.\\

To demonstrate that this protocol deterministically prepares the desired number of electron spins in the QDM, we performed a control experiment depicted in Fig.\,\ref{fig:charging_protocol}(b). The charging laser is pulsed (200\,ns duration, square pulse) and swept across the two charging voltage plateaus. The time offset between the onset of the charging laser and the beginning of the first charge voltage plateau is defined as $\Delta t$. In the readout stage ($V_G=0.56\,\mathrm{V}$), we probe the resulting charge state by setting the readout laser to be resonant with the $X^0$, $X^-$ or $X^{2-}$ transition, respectively. Typical results are presented in Fig.\ref{fig:charging_protocol}(c). When the charge laser overlaps with the first charge voltage plateau, the $X^0$ transition measured in the readout phase of the experiment quenches, to be replaced by $X^-$ at $\Delta t = -0.2\,\mathrm{\mu s}$.  This observation confirms that a single electron is prepared in the QDM during the charging phases (II and III) of the protocol. As soon as the charge laser pulse temporally overlaps with both charge plateaus (II and III) at $\Delta t = 0$, the $X^-$ signal quenches, and only emission stemming from $X^{2-}$ is observed in the readout phase. As expected, when the charge laser pulse does not overlap with either plateau, the $X^{2-}$ signal again vanishes and only $X^0$ signal is observed. As for the previously studied hole storage sample in Ref.\,\cite{bopp_quantum_2022}, we determined the charging fidelities to be above 85\,\%. A detailed calculation of the charging fidelities can be found in section \,\ref{sec:charge_fidelities} of the supplementary data.

\begin{figure*}
    \includegraphics[width=0.9\linewidth]{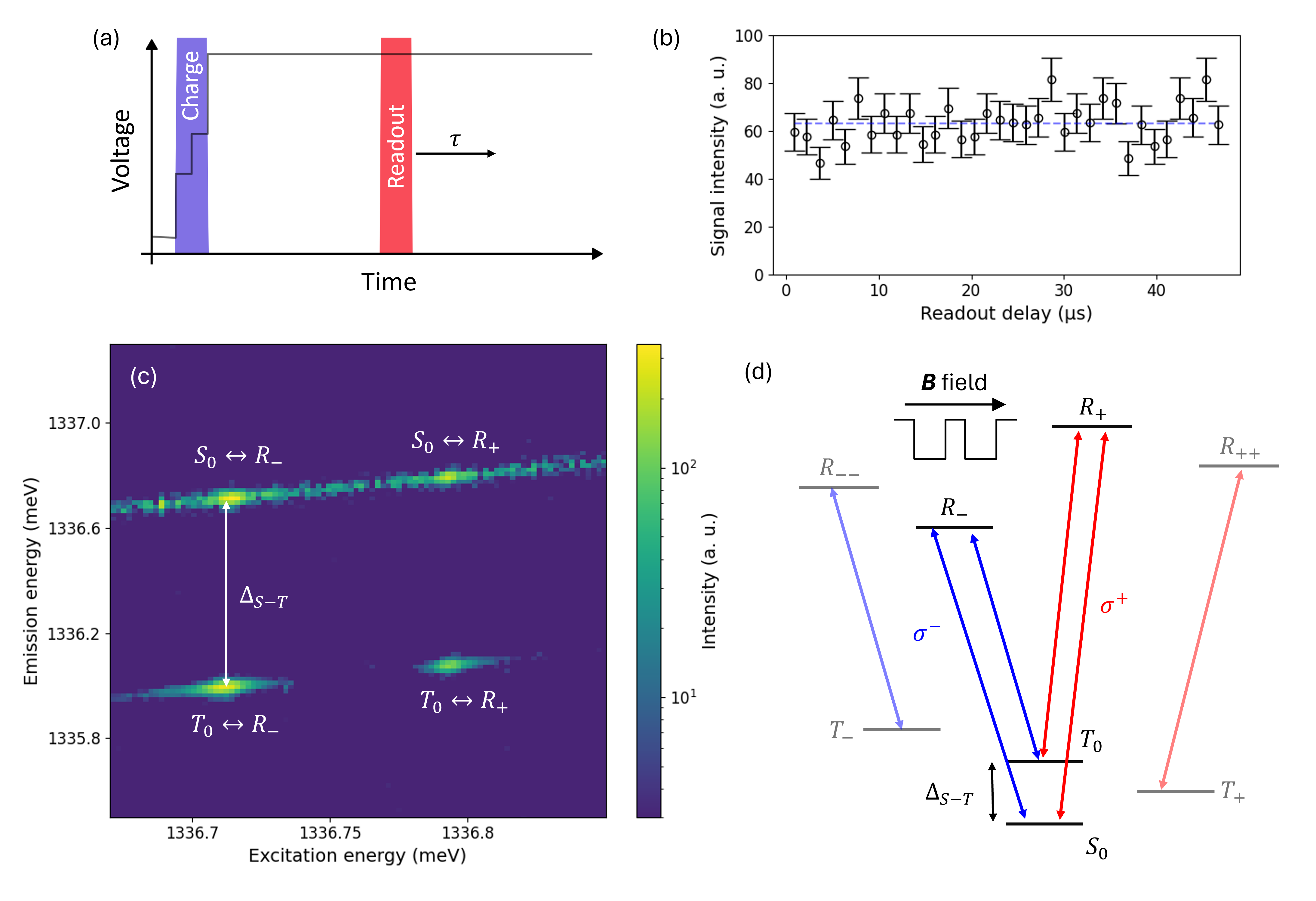}
    \caption{\label{fig:charge_storage}Measurements on hybridized two-electron state. (a) Experimental method applied to measure the charge storage time. After the QDM is charged with two electrons, a short readout pulse resonant with the $S_0\leftrightarrow X^{2-}$ transition probes the two-electron state throughout the readout phase IV of the experiment. (b) Intensity of the $X^{2-}$ signal as a function of delay of the readout pulse relative to the charging phase. (c) RF scan over the hybridized two-electron state under application of an external magnetic field of 1\,T (Faraday geometry). (d) Level scheme of the two electron states and the doubly charged exciton states used for optical excitation.}
\end{figure*}

\section{\label{sec:singlet-triplet}Singlet-triplet regime}
The previous measurements were performed at an applied gate voltage of $V_G=0.56\,\mathrm{V}$, where the two-electron ground state is (2,0) (two electrons in lower dot, zero electrons in upper dot). It is interesting to probe the situation when the orbital wavefunctions of the two electrons are hybridized over the two dots, the (1,1)-configuration and the Heisenberg coupling of the two-spins is voltage tunable. Here, the electrons form hybridized singlet and triplet states that we denote as $(1,1)_S$ and $(1,1)_T$, separated by the singlet-triplet splitting $\Delta_{S-T}$, determined by the orbital tunnel coupling between the dots\,\cite{doty_optical_2008}. This regime is also known as the ``sweet spot'', since the hybridized two-electron states are protected to the first order against electrical noise\,\cite{petta_coherent_2005, weiss_coherent_2012}. We energetically lift the degeneracy of the degenerate triplets $T_+$ and $T_-$ by applying a magnetic field in growth direction (Faraday geometry). The $S_0$ and $T_0$ states then form an optically isolated qubit that can be addressed via the (spin polarized) doubly charged exciton states $R_+=(\downarrow,\uparrow\downarrow\Uparrow)$ and $R_-=(\uparrow,\uparrow\downarrow\Downarrow)$\,\cite{doty_optical_2008}. Fig.\ref{fig:charge_storage}(d) illustrates the energy level configuration of this system. From the PL-V measurements we identify this regime at a gate voltage $V_G = 1\,\mathrm{V}$ (see also supplementary information).\\

To demonstrate that the singlet-triplet states can be optically addressed, we first show that the two electrons remain in the QDM over timescales much longer than needed for typical optical experiments.  This is a key observation since it demonstrates that optical driving does not result in charge loss due to Auger processes or photo-ionization\,\cite{lobl_radiative_2020, kurzmann_auger_2016}. We do this by performing the experimental sequence depicted schematically in Fig.\,\ref{fig:charge_storage}(a), where after charging the QDM with two electrons and moving them to the sweet spot, a short readout pulse delayed by a time $\tau$ probes the two-electron population. Due to electrons tunneling out of the QDM we expect an exponential decay of the $X^{2-}$ signal with increasing delay time. Typical results obtained from such an experiment are presented in Fig.\,\ref{fig:charge_storage}(b). Within experimental precision, we observe no decay of the $X^{2-}$ signal for times between the preparation of the $2e^-$ spin state and probing it up to $50\,\mathrm{\mu s}$. Hence, we conclude that the charge storage time is $\tau_\mathrm{storage}\gg 50\,\mathrm{\mu s}$.
We continue to investigate the optical excitation of the singlet-triplet qubit. Here, two electrons were again prepared before they were tuned to the sweet spot. A magnetic field of 1\,T is used to lift the degeneracy of the spin-polarized triplet states. The energy of the readout laser pulse is then scanned over the singlet-trion transitions. The results of this scan are shown in Fig.\,\ref{fig:charge_storage}(c). We observe two resonant transitions, $S_0\leftrightarrow R_-$ and $S_0\leftrightarrow R_+$ at 1336.713(23)\,meV and 1336.792(23)\,meV, respectively, as well as two non-resonant signals, which are the decays of the trion to the respective triplet states. These off-resonant transitions have an energy of 1335.995(23)\,meV ($T_0\leftrightarrow R_-$) and 1336.076(23)\,meV ($T_0\leftrightarrow R_+$). Away from the resonances, a background from the excitation laser is also visible, since the circular Bragg grating structure on top of the QDM limits the rejection of the scattered resonant drive laser by cross-polarized suppression between the excitation and detection channels. From the energy separation between the resonant and the non-resonant signal we directly read the singlet-triplet splitting\,$\Delta_{S-T}$. For this sample it amounts to $718(16)\,\mathrm{\mu eV}$.

\section{\label{sec:singlet-init}Singlet initialization}
\begin{figure*}
    \includegraphics[width=0.7\linewidth]{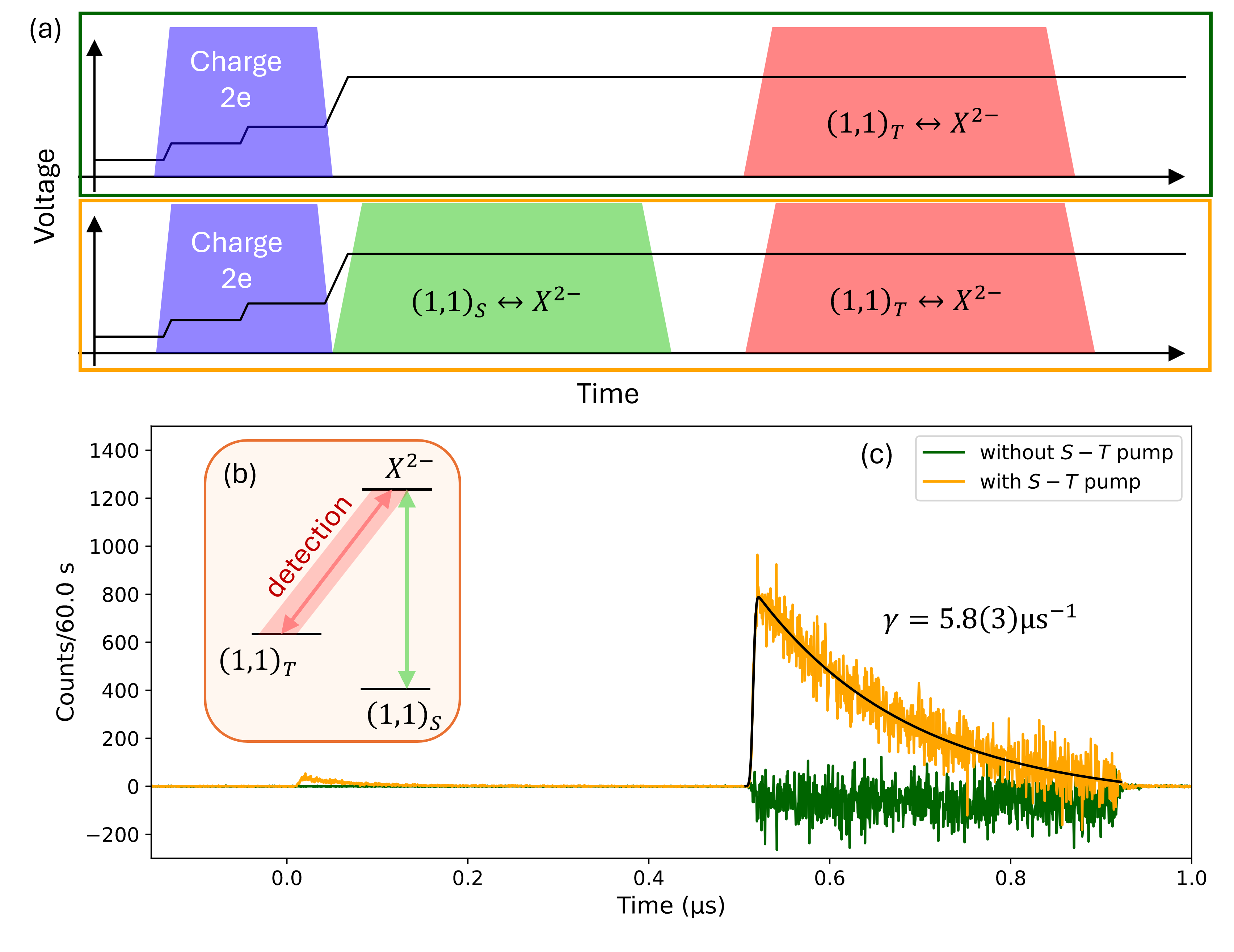}
    \caption{\label{fig:spin_init}Singlet initialization and probing of singlet-to-triplet transition in an optically controlled QDM. (a) Measurement scheme for demonstrating singlet initialization via charging. Following the preparation of two electrons, a probe pulse (red) resonant with the triplet to trion transition probes the spin population in the triplet state without (top) and with (bottom) a pump pulse (green) resonant with the singlet to trion transition prior to the triplet readout. (b) Level scheme of the two electron state and the respective direct trion used for optical excitation. The ground state is the $(1,1)_S$ state. Optically driving the $(1,1)_S\leftrightarrow X^{2-}$ transition pumps the initial singlet population into the $(1,1)_T$ state. (c) Time-resolved signal of the trion-to-triplet transition. Between 0.5\,$\mathrm{\mu s}$ and 0.9\,$\mathrm{\mu s}$ a pulse resonantly probing the triplet population is applied. The signal was recorded without (green) and with (orange) a singlet-to-trion pump pulse applied prior to readout. A strong resonant trion-to-triplet signal is only observed when pumping the two-electron population from the singlet to the triplet first, indicating that the bare two-electron state after charging is mostly in the $(1,1)_S$ configuration.}
\end{figure*}

Until now, we have demonstrated that two electron spins can be deterministically prepared in the QDM and subsequently tuned into different orbital configurations by changing the gate voltage. We continue to show that the electrons retain their singlet character even after electrically switching away from the charging spot. Here, we note that the singlet state remains the ground state of the $2e^-$ system for any given gate voltage. Therefore, spin initialization of the two-electron spin state into $(2,0)_S$ follows from thermalization without the need for additional optical spin preparation. To confirm this expectation, we performed the measurement protocol shown in Fig.\,\ref{fig:spin_init}(a). Following the preparation of two electrons, the gate voltage is switched such that the electrons are in the (1,1)-configuration. A 400\,ns cw-laser pulse (red) resonantly probes the triplet population and the recorded signal is analyzed in a time-resolved manner. The level scheme along with the excitation lasers and the detection filter setting is shown in the inset of Fig.\,\ref{fig:spin_init}(b) (see supplementary information\,\ref{sec:setup} for experimental details). The experiment is repeated in a second protocol, in which prior to probing the triplet population, another 400\,ns cw-pulse (green arrow in Fig.\,\ref{fig:spin_init}(b)) drives the singlet-to-trion transition. This optically pumps the two-electron population from the $(1,1)_S$ state to the $(1,1)_T$ state. Fig\,\ref{fig:spin_init}(b) shows the time-resolved counts of the measurement sequence with and without the $S-T$ pump pulse. Without the $S-T$ pump pulse prior to probing the triplet population, no resonant triplet signal is observed (dark green time trace). In order to observe a resonant triplet signal, one needs to first shelve the population from the $(1,1)_S$ into $(1,1)_T$. This is done by using a another cw laser whose energy is resonant with the $(1,1)_S \leftrightarrow X^{2-}$ transition (light green pulse in the measurement sequence in Fig.\,\ref{fig:spin_init}(a)). If this pump pulse is employed, then the triplet laser probes a significant resonant signal which is shown by the orange time trace in Fig.\,\ref{fig:spin_init}(b). Besides the strong resonant signal of the probe pulse, a weaker signal of trion-to-triplet decay is also visible during the pump phase. This is a clear sign that the shelving of the two-electron state from the singlet to the triplet via the $X^{2-}\rightarrow (1,1)_T$ happens during pumping.
Another observation is that the triplet signal is not constant in time, but rather shows a decaying behavior. At the chosen probe power of 2\,nW the decay rate was 5.8(3)\,$\mathrm{\mu s^{-1}}$. This observation is an indication of spin pumping occuring: The resonant triplet laser optically pumps the signal back to the singlet state. In the next section we continue to explore spin pumping and the relaxation between singlet and triplet states further.\\
We determine a lower bound for the spin initialization fidelity $F_\mathrm{init}$ from the charging process by comparing the signal observed during the triplet measurement, both with and without $S-T$ pump phase of the measurement protocol. The resonant triplet signal without the prior pump pulse directly measures the initial triplet population, which is vanishingly small. The limiting factor in this estimation stems from the subtraction of residual resonant laser, which leads to substantial background noise. The triplet signal is buried by this noise, such that we estimate the maximum counts hidden within the laser background by measuring the 95\,\% confidence interval $\sigma_{95\%}$. Multiplying this value by the on-time of the probe laser (400\,ns) allows us to calculate the total number of photon counts $I_\mathrm{no\,pump, 95\%}$ that could stem from residual triplet population without prior $S-T$ pump pulse. This value can be directly related to the total integrated counts $I_\mathrm{pump}$ of the triplet signal with the prior $S-T$ pump pulse active. We estimate the absolute lower bound for the fidelity $F_\mathrm{init} = 1 - \frac{I_\mathrm{no\,pump, 95\%}}{I_\mathrm{pump}} \gg 61.2\,\%$.
As mentioned, this low value is significantly underestimated since it is based on the unlikely assumption that a constant triplet signal may be buried within the laser background. Most likely, the actual initialization fidelity is much larger.\\

The measurements suggest that electrical switching does not change the singlet character of the initially prepared $(2,0)_{S_0}$-state. This is an expected result, since triplet and singlet electron states (in the absence of holes) are only weakly coupled by nuclear or external magnetic fields\,\cite{wilksen_gate-based_2024}. Mixing with with higher energy singlet states remains a possibility, however even at the singlet anticrossing point, the energy gap between two singlet states is $\sim 2\,\mathrm{meV}$, or $\sim 484\,\mathrm{GHz}$. The adiabatic theorem\,\cite{born_beweis_1928, kato_adiabatic_1950} states that in order for singlet mixing to happen during electric switching, the switching speed needs to be in the same order of magnitude as the energy difference. The maximum bandwidth of the waveform generator is 100\,MHz and therefore over three orders of magnitude lower than the required speed to create singlet mixing.

\section{\label{sec:spin_pumping} Optical pumping of one and two electron states and triplet-singlet-relaxation}

\begin{figure*}
    \includegraphics[width=0.9\linewidth]{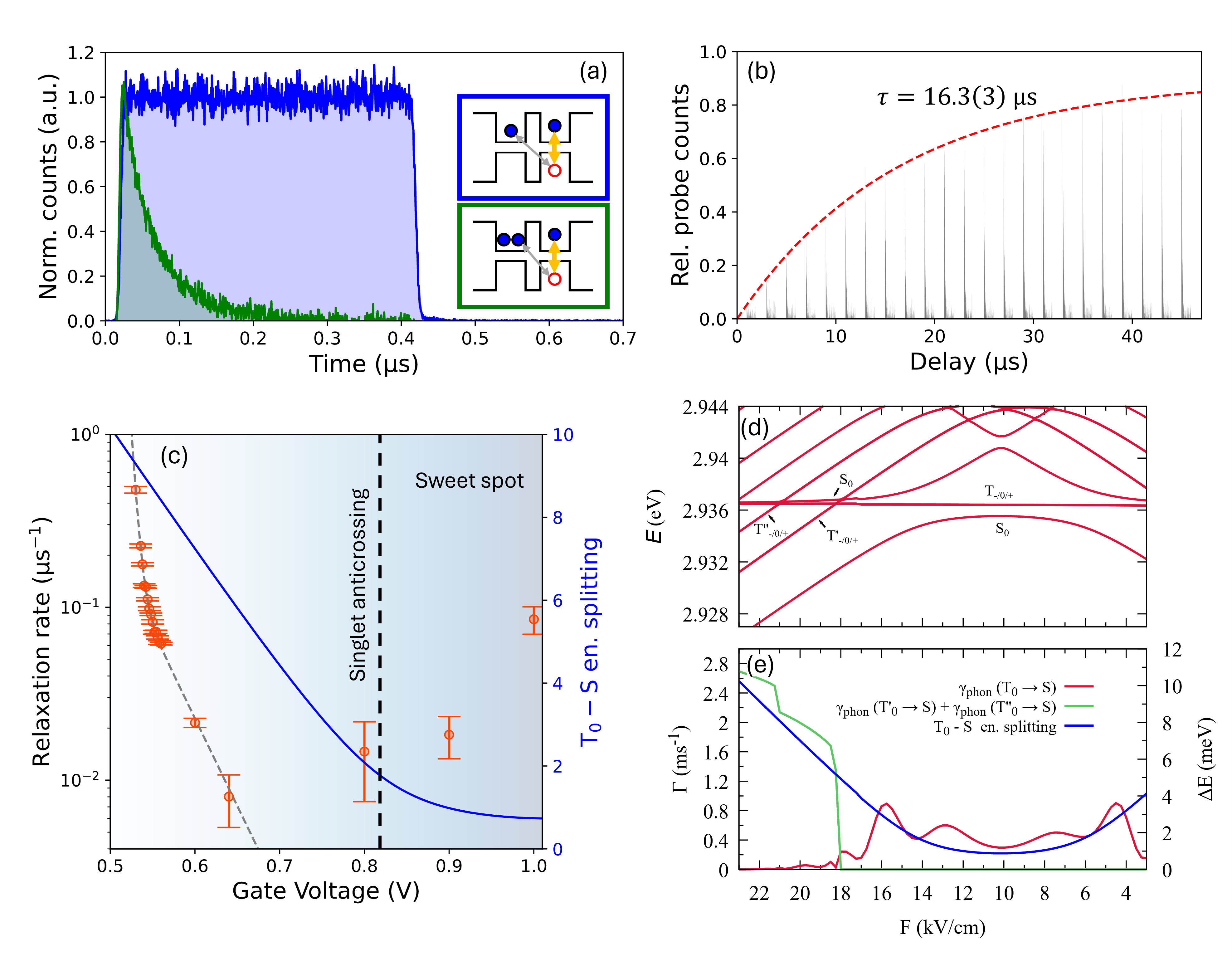}
    \caption{\label{fig:spin_pumping}Orbital pumping and relaxation of one- and two-electron states. (a) Optical driving of the one-electron (1,0) and two-electron (2,0) systems demonstrating Pauli blockade, and optical pumping of (2,0) driven via the direct $X^{2-}$-transition. The insets depict the energy level scheme of the singly and doubly charged QDM subject to resonant optical excitation of the upper dot. (b) Typical results of pump-probe measurements consisting of optical driving of (2,0) via the direct $X^{2-}$-transition to shelve population in (1,1) before a delay without optical driving and a re-pump pulse to test relaxation from (1,1)$\rightarrow$(2,0). (c) Orbital relaxation time measured from cw-pump-probe experiments as a function of the gate voltage.  The anticrossing of $(2,0)$ and $(1,1)_{S}$ singlet states is marked by the vertical dashed line and the S-T sweet spot is close to 1.0V.  (d) Calculated field dependence of single and triplet states from the 8 band $k \cdot p$ modeling, revealing sweet spot around ${10}$kV/cm. Note the higher energy triplet state shifting below the excited $(1,1)_{T_0}$ triplet at $F\geq18$ kV/cm. (e) Calculated singlet-triplet-relaxation rates as function of the electric field.}
\end{figure*}
we continue by investigating optical pumping and relaxation of one and two electron states in the QDM. In the previous section it was already shown that under resonant excitation of the two-electron state, the state population can be optically pumped from a singlet to a triplet and vice versa, without a magnetic field. Our charge tunable device facilitates measurement of the spin relaxation of the coupled electron system over a large range of orbital energy detunings to directly compare the situation to a single electron in the QDM. For two electrons, we uncover some intriguing and surprisingly slow dynamics.\\

We first compare optical pumping of the single electron state and the two electron state in a voltage-regime, where the electron(s) orbital wavefunctions are mostly confined in the lower dot, i.e. the (1,0) configuration and (2,0) configuration, respectively. By optically exciting a singly (doubly) charged exciton in the upper dot, the system can either decay directly to its original state, or indirectly with rate $\gamma_\mathrm{P}$ with the electron being effectively pumped into the upper dot. An electron in the upper dot will eventually relax back to the lower energy $(1,0)$-state with rate $\gamma_\mathrm{rel}$. The situation is depicted by the inset in Fig.\,\ref{fig:spin_pumping}(a) for both the single and two-spin cases. The blue trace in Fig.\,\ref{fig:spin_pumping}(a) shows the time-trace of the single-electron case and compares with the two-electron (green trace) upon resonant driving. A clear qualitative difference between the single-electron case and the two-electron case is visible: While the resonant signal remains constant when driving a single electron via the trion, the driving the two electrons case via the doubly charged exciton reveals an exponential decay over nanosecond timescales. This is a clear indication for optical spin pumping. The fact that no such a decay is observed for a single electron can be understood by phonon-assisted tunneling: While the indirect transition does indeed occur and is also observed in luminescence, time-resolved studies have shown that inelastic interdot tunneling can occur over timescales ranging from ps $\rightarrow$ 100 ps ,\cite{muller_electrical_2012}. This is much faster than the timescale of optical pumping such that optical pumping is inhibited.\\

The fact that this is not observable in the two-electron case suggests that the relaxation from the (1,1)-state back to the (2,0)-state is inhibited by Pauli blocking and that the relaxation time is much longer than the timescale for optical pumping. In \,\ref{sec:selection_rules} of the supplementary information, we present detailed calculations and analysis of the selection rules that lead to the spin pumping mechanism. The result is that for continuous pumping, the two-electron population becomes shelved from the $(2,0)_S$-state into the $(1,1)_{T_0}$, which is a metastable configuration with regard to the interdot-tunneling timescales. We quantify the stability of the $(1,1)_{T_0}$-state by directly measuring the relaxation time in a pump-probe-like experiment, where we first use a resonant cw-pulse to pump the $2e$ population from $(2,0)_S$ to $(1,1)_{T_0}$ and after some delay time we apply a second, identical pulse to probe how much of the population has relaxed back to $(2,0)$. This directly probes the triplet-singlet-relaxation rate. Fig.\,\ref{fig:spin_pumping}(b) shows the probe histograms as a function of the delay time at $V_G=0.56\,\mathrm{V}$. From the fit we can see that the relaxation takes place on a timescale of $\tau_\mathrm{rel}=16.3(\pm 0.3)\,\mathrm{\mu s}$ (corresponding to a relaxation rate of $\gamma_\mathrm{rel}=0.061(2)\,\mathrm{\mu s}^{-1}$) and is thus significantly longer than the picosecond-timescale expected for a single electron.\\

We measure the triplet-singlet relaxation rate over a larger voltage domain and thus different orbital configurations, from a regime where the $(2,0)$ configuration is the ground state, all the way to the hybridization regime at 1\,V. The measured relaxation rates are shown in Fig.\,\ref{fig:spin_pumping}(c). At low voltages and thus large electric field, the orbital character of the singlet ground state is mostly $(2,0)$ and the relaxation rate drops exponentially as a function of voltage. We thus model the voltage-dependent relaxation rate as a function
\begin{equation}
\gamma_\mathrm{rel}(V) = \gamma_\mathrm{rel, 0}\exp(-\kappa (V-V_0))    
\end{equation}
with some reference voltage $V_0$ and reference relaxation rate $\gamma_\mathrm{rel, 0}=\gamma_\mathrm{rel}(V_0)$. We find $\kappa = 187.57(10.71)\,\mathrm{V}^{-1}$ until 0.54\,V. After that, the exponential rate at which the relaxation rate decreases is lower, at $\kappa = 23.33(3.89)\,\mathrm{V}^{-1}$ which continues until at least $V_G=0.65\,\mathrm{V}$. The fact that the triplet-singlet-relaxation rate depends so strongly on the gate voltage signifies a strong electric field dependence, i.e. a strong dependence on the energy splitting of the orbital states involved. Furthermore, the two different timescales observed change rather abruptly in two different voltage domains hints towards different primary decay mechanisms being present in the two regimes. As a reminder, until a gate voltage of 0.75\,V the orbital character of the singlet ground state is predominantly $(2,0)$, while the triplet is always $(1,1)$. For $V_G>0.65\,\mathrm{V}$ the relaxation rate becomes too large for us to be able experimentally probe it. However, towards the singlet hybridization regime, for $V_G\mathrm{>0.8\,V}$, we observe an increase in the relaxation rate, reaching a value of $\gamma_\mathrm{rel}=0.085(182)\,\mathrm{\mu s}^{-1}$ in the sweet spot at $V_G = 1\,\mathrm{V}$. This increase is accompanied by the change of the orbital character of the singlet ground state from $(2,0)_S$ to $(1,1)_S$. This change is indicated in Fig.\,\ref{fig:spin_pumping}(e) by the background from white to blue. At the same time, the energy splitting between the $(1,1)_T$ state and the singlet ground state changing from being linearly dependent on the gate voltage to flattening out towards the sweet spot and reaching its minimum of 718\,$\mathrm{\mu eV}$. In Fig.\,\ref{fig:spin_pumping}(e) the blue curve shows the triplet-singlet energy splitting as a function of the gate voltage. The relaxation between two-electron states is much longer compared to the single electron since relaxation from one of the triplet states inherently requires a spin flip, since the spin projection of the two-electron system must change from $M_S$=1 to 0. Hence, spin-orbit mediated spin relaxation must take place.\\

Recently, a theoretical analysis of triplet-singlet relaxation in QDMs was presented by some of us\cite{gawarecki_phonon-assisted_2021}. The approach is based on eight-band $k \cdot p$ theory and takes into account spin-orbit-coupling that mixes pure spin states, thereby opening up channels for phonon mediated relaxation~\cite{Khaetskii2000,Khaetskii2001}. Here, we employed this framework adapted to the size, shape and composition of the QDM studied. As discussed in more detail in section ~\ref{sec:theory_kp} of the supplementary information, the size, shape and In-composition profile of the QDM was adjusted to fit our optical spectroscopy data. The electric field dependence of the lowest energy two-electron states were then computed by diagonalization of the Coulomb Hamiltonian using the configuration interaction method \cite{Gawarecki2018a, Lienhart2025}. 
The calculated energies of the two-spin eigenstates are presented in Fig.\,\ref{fig:spin_pumping}(d) for the case with $B=0.1$~T applied in Faraday geometry. This was chosen to be larger than the typical Overhauser fields for electrons in self-assembled dots, justifying neglecting hyperfine mediated relaxation in our model~\cite{urbaszek_nuclear_2013}. We then computed the pure phonon-assisted spin-orbit mediated relaxation rates between the two-electron energy levels including electron-phonon interaction mediated by deformation potential and piezoelectric couplings~\cite{Yu2005,Grodecka2005a,Krzykowski2020} (see supplemental, section ~\ref{sec:theory_kp}).    

The two-spin eigenstates and the one- and two-phonon relaxation rates, presented in Fig.\,\ref{fig:spin_pumping}(d) and (e), respectively, as a function of electric field. Focusing first on the spin and orbital configurations at higher electric field (e.g. $F \approx 22$~kV/cm), we note that the lowest energy state is the spin singlet localized in the lower QD $\ket{S_{\ell}}$. The next two energetically higher lying states, labeled $\ket{T'_{-/0/+}}$ and $\ket{T''_{-/0/+}}$, have triplet character and are each admixtures of $s$ and $p$ orbitals in the lower dot (see supplemental, section ~\ref{sec:theory_kp}). For orientation, the blue curve on Fig.\,\ref{fig:spin_pumping}(e) shows the field dependent S-T splitting between $\ket{T_{-/0/+}}$ and $\ket{S_{0}}$. In the vicinity of the sweet spot ($F\approx 10$~kV/cm) we calculate a S-T splitting of $870\,\mu$eV, close to the experimentally measured value ($700\,\mu$eV), confirming the general validity of our eight-band $k \cdot p$ model. 

The field dependent one-phonon (red curve) and two-phonon (green curve) relaxation rates are presented in Fig.\,\ref{fig:spin_pumping}(e).  For $F>10$~kV/cm, both $\ket{T'_{-/0/+}}$ and $\ket{T''_{-/0/+}}$ progressively shift energetically below $\ket{S_l}$. This is accompanied by a rapid increase in the two-phonon relaxation rate (green curve), an observation qualitatively consistent with rapid increase in the two-spin relaxation rate observed experimentally for gate voltages below $0.54$~V (Fig.\,\ref{fig:spin_pumping}(c)). Fundamentally, the rapid increase arises from transitions via intermediate triplet states involving $p$-shell orbitals in the lower dot. 

%
The calculated total phonon mediated relaxation rates are $\sim50\times$ smaller than we observe in experiment. However, we note that our measurements were performed at zero magnetic field, where phonon-assisted hyperfine spin-flips may also contribute significantly to the relaxation. Until now phonon-assisted hyperfine spin-flips have so far only been theoretically studied in QDMs for single electron doped dots\,\cite{karwat_phonon-assisted_2021}.   Moreover, relaxation rates are likely to be strongly impacted by the precise admixture of $s$- and $p$-orbitals in the triplet and lower singlet states. This depends on the precise orbital level structure of QDMs, likely to be a complex function of size, shape and In-composition.  As such, quantitatively taking into account all different states and relaxation pathways, particularly in the two-particle case, is expected to be non-trivial. We consider the quantitative agreement between experiment and theory to be satisfactory and the qualitatively similar electric field dependence provides good evidence that one- and two-phonon mediated processes are indeed responsible for relaxation.  Further investigations are required to fully understand the role of hyperfine coupling in the QDMs.

\section{\label{sec:conclusion} Conclusion and outlook}
In summary, we demonstrated the sequential optical charging of a single QDM with two electron spins while simultaneously maintaining the ability to widely tune orbital couplings using static electric fields and optically drive the system for quantum light generation. We optically prepared one- and two-spin states, initialized via optical pumping and explored orbital and spin relaxation dynamics for one and two-spin states as a function of the energy detuning and hybridization of orbital states. For two-spin states, remarkably long S-T relaxation times were observed extending beyond $\sim 100\mu s$ with strong dependence on the relative energy of ground and excited two-spin states.  Semi quantitative agreement was observed with $\mathbf{k \cdot p}$ calculations of phonon-mediated spin-relaxation

Our work demonstrates that charge storage devices can be used to deterministically prepare two electrons in a quantum dot molecule with hybridized orbital states. This forms the basis of various quantum information applications, such as the generation of two-dimensional cluster states. The ability to tune the orbital wavefunction configuration of the two-electron-state without changing the charge state furthermore opens up possibilities to use electrical switching to perform gate operations\,\cite{wilksen_gate-based_2024}.

\section*{Acknowledgments}
JJF and KM gratefully acknowledge funding by the Bavarian Hightech Agenda within the Munich Quantum Valley. We acknowledge the BMFTR for financial support via the QR.N consortium via sub-projects FKZ 16KIS2197 (JJF), BMFTR via QR. N project 16KIS2200 and 16KIS2203, BMFTR via QUANTERA EQSOTIC project 16KIS2061, BMFTR via EUROSTARS QTRAIN project 13N17328, as well as the DFG excellence cluster ML4Q project EXC 2004/1 (AL), 16KIS2193 (SR), 16KIS2203 (CG) and 16KIS2206 (DR) and the Deutsche Forschungsgemeinschaft (DFG, German Research Foundation) under Germany's Excellence Strategy -- EXC-2111 -- 390814868 and project CNLG (MU 4215/4-1) (KM). KG and JJF acknowledge the DAAD-NAWA for financial support via the center-to-center exchange Grant No. 57754510. The project is co-financed by the Polish National Agency for Academic Exchange.

\bibliographystyle{apsrev4-2}
\bibliography{bibliography2,KG_bibliography}

\clearpage
\newpage
\onecolumngrid
\appendix

\section*{Supplementary Information for 'Optically probing one and two electron 
orbital and spin states in an electrically tunable quantum dot molecule'}
\addcontentsline{toc}{section}{Supplementary Information}
\makeatletter
\def\appendixname{}
\renewcommand{\thesection}{S\arabic{section}}
\renewcommand{\thefigure}{S\arabic{figure}}
\renewcommand{\thetable}{S.\Roman{table}}
\long\def\@hangfrom@section#1#2#3{\@hangfrom{#1#2}\MakeTextUppercase{#3}}
\def\@sectioncntformat#1{\csname the#1\endcsname.\quad}
\setcounter{section}{0}
\setcounter{figure}{0}
\renewcommand{\theequation}{S\arabic{section}.\arabic{equation}}
\setcounter{equation}{0}
\makeatother

\section{\label{sec:sample_design}Sample design}
\begin{figure}[ht]
    \includegraphics[width=0.9\linewidth]{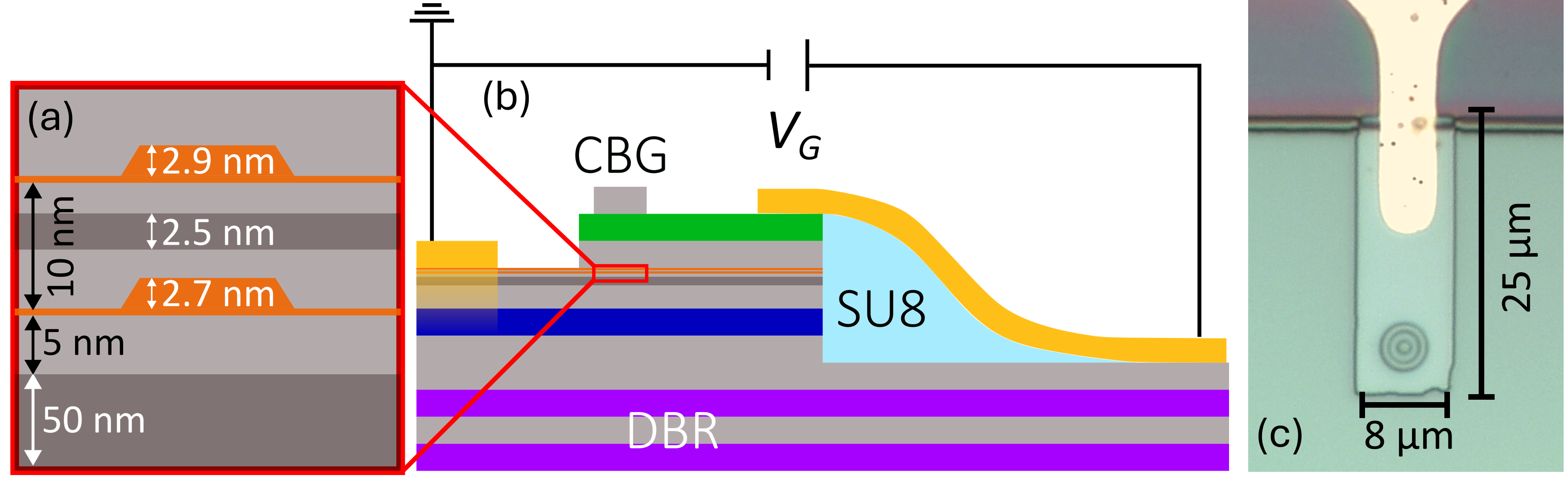}
    \caption{\label{fig:sample_design}Sample design (a) Schematic close-up to the two tunnel-coupled quantum dots with the relevant measures. Light-gray shaded areas represent the GaAs-matrix, while dark-gray shaded areas represent $\mathrm{Al_{0.34}Ga_{0.66}As}$-barriers. The orange-shaded area shows the stacked quantum dots with the wetting layers below them. (b) Schematic cross section of the full layer structure and device. The green layer is the $p$-doped layer, while the dark blue layer is the $n$-doped layer. The $n$-contact on the left does not directly touch the $n$-doped layer, but is thermally annealed to it. DBR: Distributed Bragg reflector, CBG: Circular Bragg grating. (c) Microscope image of the fabricated device.}
\end{figure}

The measurements were performed on two vertically stacked InAs quantum dots embedded in a GaAs matrix. The quantum dot heterostructure is schematically shown in Fig.\,\ref{fig:sample_design}(a). It was grown molecular beam epitaxy in the Stranski-Krastanov growth mode, where the lower and upper dots were In-flushed to a height of 2.7\,nm and 2.9\,nm, respectively. The total separation between the two dot layers is 10\,nm. The separating layer mostly consists of GaAs, but 2.5\,nm of it are made from $\mathrm{Al_{0.34}Ga_{0.66}As}$ to further reduce the tunnel coupling strength between the stacked dots. 5\,nm below the lower quantum dot layer there is a 50\,nm $\mathrm{Al_{0.34}Ga_{0.66}As}$ layer that serves as a tunneling barrier for electrons trapped in the quantum dots. The whole structure is sandwiched by a $p$-doped layer (top) and an $n$-doped layer (bottom) for electrical contacting. To increase the photonic out-coupling efficiency, a distributed Bragg reflector (DBR) is located at the bottom of the layer structure, combined with a circular Bragg grating (CBG) that is deterministically etched on top of individual QDMs using in-situ electron beam lithography at 10\,K\,\cite{schall_bright_2021}. These are then embedded into a p-i-n diode, where the QDMs are contacted individually by metallic front contact. An electric field is applied to the (individual) QDM by applying a gate voltage $V_G$ between the two contacts. The $p$-layer of the diode is kept small ($25\,\mathrm{\mu m}\times8\,\mathrm{\mu m}$) to keep its capacitance low and and therefore electrical switching times short. To make the contacts smooth, an insulating bridge, made from SU-8 resist, was fabricated first. A schematic cross-section of the device is shown in Fig.\,\ref{fig:sample_design}(b), while Fig.\,\ref{fig:sample_design}(c) shows a microscope image of a real fabricated device.

\section{\label{sec:charge_states}Identification of different charge states}
In this section we discuss the the identification of the different charged complexes observed in the voltage-dependent photoluminescence spectra presented in section\,\ref{sec:plv} of the main text. We use phenomenological few-spin Hamiltonians that take into account Coulomb interactions between the particles, as well as coherent tunnel coupling, which govern the voltage-dependence of the observed PL lines. Effects that cannot be resolved by this PL method, e.g. the electron-hole-exchange interaction, are omitted here. The identification of the charged complexes from the PL-V assists us finding the resonant energies used in the experiments in sections\,\ref{sec:charging}, \ref{sec:singlet-triplet}, \ref{sec:singlet-init} and \ref{sec:spin_pumping}. The discussions closely follows the models presented in\,\cite{doty_optical_2008}.
\subsection{Neutral exciton}
\begin{figure}
    \includegraphics[width=0.7\linewidth]{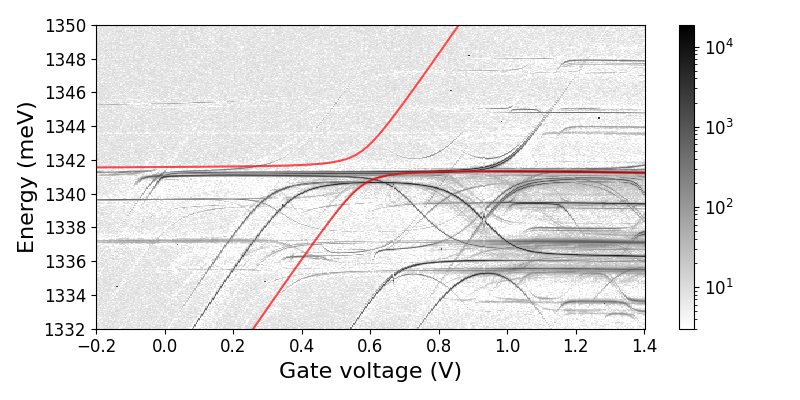}
    \caption{\label{fig:SI_X0_PLV}Photoluminescence spectrum overlaid with the $X^0$ energies obtained from the model explained in the main text.}
\end{figure}

The voltage-dependent photoluminescence (PL-V) spectra show different charged excitonic complexes. From this data, we can identify the neutral exciton ($X^0$), the singly charged exciton ($X^-$) and the doubly charged exciton ($X^{2-}$) by modeling the different direct and indirect charge configurations\,\cite{doty_optical_2008, schall_bright_2021}. Using the notation explained in the main text, we can write the neutral exciton with the basis states
\begin{equation}\label{eq:X0}
    \begin{pmatrix}
    1 & 0 \\
    0 & 1
    \end{pmatrix},\quad \begin{pmatrix}
    0 & 1 \\
    0 & 1
    \end{pmatrix}.
\end{equation}
In this basis we can express the Hamiltonian of the $X^0$ as (omitting the electron-hole exchange interaction)
\begin{equation}
    \hat{H}_{X^0} = E_{X^0} \cdot \hat{I} + \begin{pmatrix} 
    \Gamma_{X^0} - f & t \\
    t & 0
    \end{pmatrix} + \begin{pmatrix}
    pF + \alpha F^2 & 0 \\
    0 & pF + \alpha F^2
    \end{pmatrix},
\end{equation}
where $E_{X^0}$ is the emission energy of the direct exciton with no electric field applied, $\Gamma_{X^0}$ is the energy mismatch between the electron levels in the upper and the lower dot, including Coulomb interaction between the electron and the hole, $t$ is the tunnel coupling strength of the electron levels and $f=edF$ with $e$ the elementary charge, $d$ the distance between the two dots and $F$ the applied electric field. The last term in \eqref{eq:X0} contains the terms related to the quantum confined Stark effect. Here, $p$ denotes the intrinsic dipole moment and $\alpha$ the polarizability. The electric field $F$ can be obtained from the applied gate voltage $V_G$ using
\begin{equation}
    F=\frac{V_\mathrm{bi}-V_G}{D}
\end{equation}
where $V_\mathrm{bi}$ is the built-in voltage of the diode and $D$ is the width of the diode's intrinsic region. For a p-i-n diode in GaAs we can estimate we can estimate the built-in voltage to be $V_\mathrm{bi}=1.51(1)\,\mathrm{V}$. The width of the intrinsic region $D=365\,\mathrm{nm}$ is known from the sample growth. Diagonalizing the Hamiltonian \eqref{eq:X0} will give us the eigenenergies of $X^0$ as a function of the applied gate voltage $V_G$. Since the neutral exciton recombines to the crystal ground state, these eigenenergies correspond to the emission energies observed in the PL-V measurement. Therefore, the PL-V spectrum allows us to determine the parameters used in this model. Fig.\,\ref{fig:SI_X0_PLV} shows the PL-V spectrum overlaid with the neutral exciton energy we obtain from our model. The determined parameters are listed in Table\,\ref{tab_X0}.

\begin{table}[ht]
    \caption{Neutral exciton parameters determined from voltage-dependent photoluminescence.}
    \label{tab_X0}
    \begin{tabular}{c|c}
         Parameter & Value \\\hline
         $E_{X^0}$ & 1341.3(1)\,meV\\
         $d$ & 10.9(1)\,nm\\
         $t$ & 1.0(1)\,meV\\
         $\Gamma_{X^0}$ & 27.8(1)\,meV\\
         $p$ & $0.14(5)\,e\cdot\mathrm{nm}$\\
         $\alpha$ & $-18(6)\,e\cdot\mathrm{nm}^2/\mathrm{V}$
    \end{tabular}
\end{table}

\subsection{Singly charged exciton}
\begin{figure}
    \includegraphics[width=0.7\linewidth]{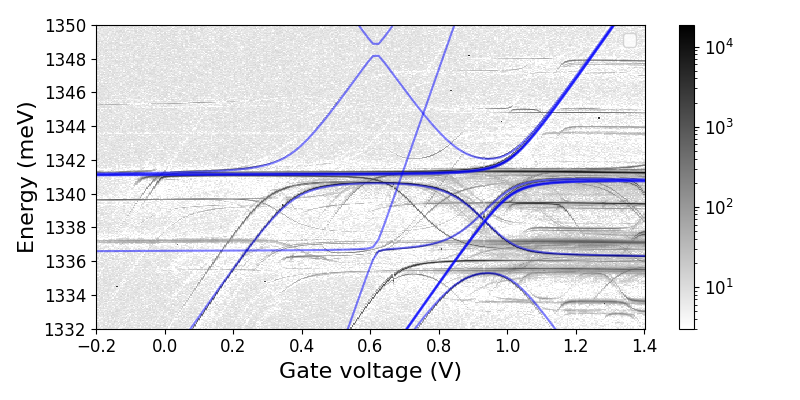}
    \caption{\label{fig:SI_Xm_PLV}Photoluminescence spectrum overlaid with the $X^-$ energies obtained from the model explained in the main text.}
\end{figure}

In a similar fashion we obtain the emission energies for the singly charged exciton. It consists of two electrons and one hole, where the hole is again located in the upper dot. Since there are now two electrons in the QDM, spin effects need to be taken into account, as well. Expanding the notation introduced in the previous section, we can express the Hamiltonian in the following basis\,\cite{scheibner_spin_2007}:
\begin{equation}
    \begin{pmatrix}
    \uparrow\downarrow & 0 \\
    0 & \Uparrow
    \end{pmatrix}, \;
    \begin{pmatrix}
    \uparrow & \downarrow \\
    0 & \Uparrow
    \end{pmatrix}_S, \;
    \begin{pmatrix}
    \uparrow & \downarrow \\
    0 & \Uparrow
    \end{pmatrix}_{T_0}, \;
    \begin{pmatrix}
    \uparrow & \uparrow \\
    0 & \Uparrow
    \end{pmatrix}_{T_+}, \;
    \begin{pmatrix}
    \downarrow & \downarrow \\
    0 & \Uparrow
    \end{pmatrix}_{T_-}, \;
    \begin{pmatrix}
    0 & \uparrow\downarrow \\
    0 & \Uparrow
    \end{pmatrix}.\label{eq:Xm_basis}
\end{equation}
The suffices indicate whether the spins are configured a singlet ($S$) or triplet ($T$). In this basis we can express the Hamiltonian of the tunnel coupled system as
\begin{align}
\hat{H}_{X^{-}} =& E_{X^-}\cdot\hat{I} + (pF+\alpha F^2)\cdot \hat{I}\notag\\&+\begin{pmatrix}
\Gamma_{ll} - 2\cdot f & \sqrt{2}t & 0 & 0 & 0 & 0 \\
\sqrt{2}t & \Gamma_{lu} - f & J_{eh} & 0 & 0 & \sqrt{2}t \\
0 & J_{eh} & \Gamma_{lu} - f & 0 & 0 & 0 \\
0 & 0 & 0 & \Gamma_{lu} - f + J_{eh} & 0 & 0 \\
0 & 0 & 0 & 0 & \Gamma_{lu} - f - J_{eh}& 0 \\
0 & \sqrt{2}t & 0 & 0 & 0 & 0
\end{pmatrix},\label{eq:X-_Hamiltonian}
\end{align}
where $E_{X^-}$ is the energy of the direct trion in the upper dot $\begin{pmatrix}0 & 2 \\ 0 & 1\end{pmatrix}$, the $\Gamma_{ij}$, $i,j\in\{l,u\}$ are energy corrections when finding the electrons in the lower ($l$) or upper ($u$) that stem from the energy mismatch of the two dots, as well as different Coulomb energies. $J_{eh}$ is the electron-hole exchange interaction, which is non-negligible for the coupled two-electron system with a localized hole.\\
The single electron basis that we use is (omitting spin again)
\begin{equation}
    \begin{pmatrix}
    1 & 0 \\
    0 & 0
    \end{pmatrix}, \;
    \begin{pmatrix}
    0 & 1 \\
    0 & 0
    \end{pmatrix}.
\end{equation}
In this basis, the Hamiltonian reads
\begin{equation}
    \hat{H}_{e^{-}} = \begin{pmatrix}
    \Gamma_{e^-} - f & t \\
    t & 0
    \end{pmatrix}\label{eq:Hamiltonian_e-}
\end{equation}
where $\Gamma_{e^-}$ is the energy mismatch between the single confined electron levels of the upper and lower dot.
We obtain the emission energies of the singly charged exciton by diagonalizing the trion Hamiltonian \eqref{eq:X-_Hamiltonian} and the single electron Hamiltonian \eqref{eq:Hamiltonian_e-} and then subtracting the single-electron eigenenergies from the trion eigenenergies. Again, this method allows us to determine the parameters of the model by comparing the transition energies obtained from the model to the emission energies of the PL-V spectrum. The PL-V along with the modeled transition energies is shown in Fig.\,\ref{fig:SI_Xm_PLV}. The found parameters are listed in Table\,\ref{tab_Xm}. Parameters, that are the same for each charged complex and were already listed in Table\,\ref{tab_X0}, are not repeated here.

\begin{table}[ht]
    \caption{Singly charged exciton parameters determined from voltage-dependent photoluminescence.}
    \label{tab_Xm}
    \begin{tabular}{c|c}
         Parameter & Value \\\hline
         $E_{X^-}$ & 1336.3(1)\,meV\\
         $\Gamma_{ll}$ & 52.9(1)\,meV\\
         $\Gamma_{ul}$ & 19.4(1)\,meV\\
         $J_{eh}$ & 0.1(1)\,meV\\
         $\Gamma_{e^-}$ & 14.8(1)\,meV
    \end{tabular}
\end{table}

\subsection{Doubly charged exciton}
\begin{figure}
    \includegraphics[width=0.9\linewidth]{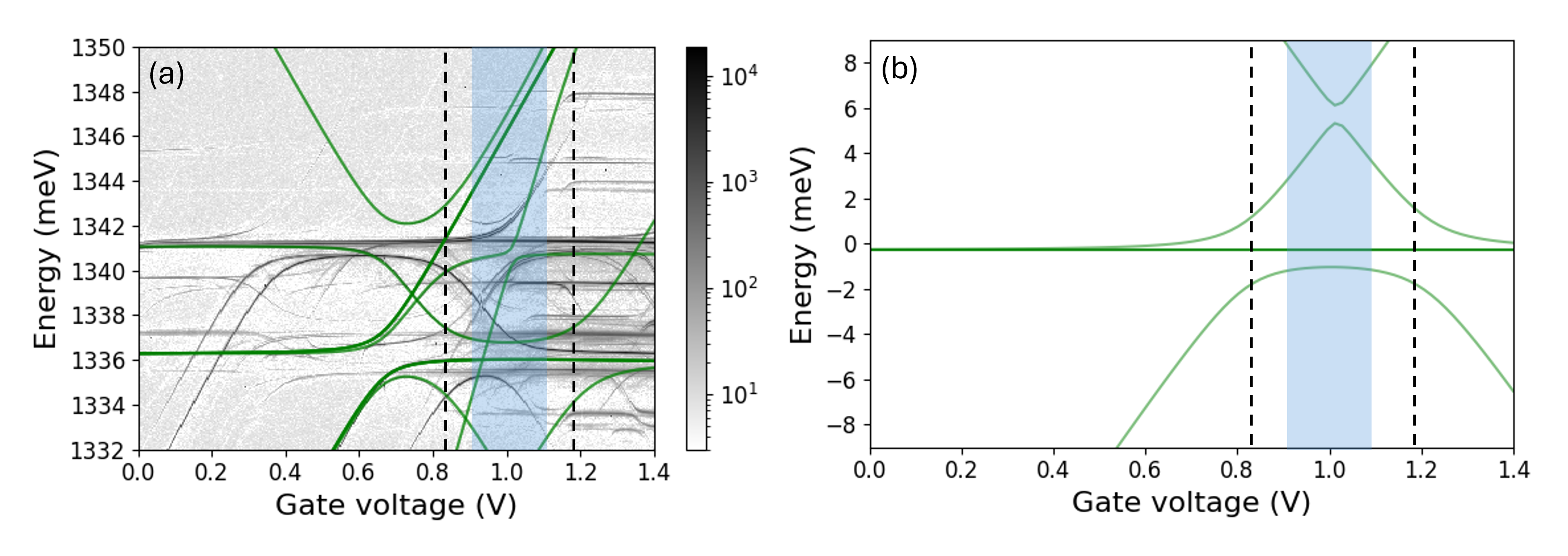}
    \caption{\label{fig:SI_X2m_PLV}(a) Voltage-dependent photoluminescence spectrum overlaid with the $X^{2-}$ energies obtained from the model explained in the main text. (b) Voltage-dependent two-electron-eigenenergies determined from diagonalizing the Hamiltonian\,\eqref{eq:e2-_Hamiltonian} and using the parameters determined from fitting the transition energies with the model explained in the main text. In both figures, the ``sweet spot'' (hybridization regime of the two electrons) is shaded in light blue.}
\end{figure}

From the emission energy we can also identify the doubly charged exciton (three electrons, one hole in upper dot). For this, we consider the doubly charged exciton in the basis
\begin{equation}
    \begin{pmatrix}
    2 & 1 \\
    0 & 1
    \end{pmatrix}, \;
    \begin{pmatrix}
    1 & 2 \\
    0 & 1
    \end{pmatrix}.
\end{equation}
We note once more that here the electron-hole-exchange interaction was omitted for simplicity. It section\,\ref{sec:selection_rules} it will become relevant. The Hamiltonian in this basis reads
\begin{equation}
    \hat{H}_{X^{2-}} = E_{X^{2-}} \cdot \hat{I} + (pF + \alpha F^2) \cdot \hat{I} + \begin{pmatrix}
    \Gamma_{X^{2-}} - f & t \\
    t & 0
    \end{pmatrix}.\label{eq:X2m_Hamiltonian}
\end{equation}
where $E_{X^{2-}}$ is the energy of the direct doubly charged exciton in the configuration $\begin{pmatrix}1 & 2 \\ 0 & 1\end{pmatrix}$.
We can further write the Hamiltonian of the two-electron system in the two-electron basis (identical to the system in \eqref{eq:Xm_basis} except that there is no hole in the upper dot):
\begin{align}
\hat{H}_{2e^{-}} = \begin{pmatrix}
\Gamma_{ll} - f & \sqrt{2}t & 0 & 0 & 0 & 0 \\
\sqrt{2}t & 0 & 0 & 0 & 0 & \sqrt{2}t \\
0 & 0 & 0 & 0 & 0 & 0 \\
0 & 0 & 0 & 0 & 0 & 0 \\
0 & 0 & 0 & 0 & 0 & 0 \\
0 & \sqrt{2}t & 0 & 0 & 0 & \Gamma_{uu}+f
\end{pmatrix}.\label{eq:e2-_Hamiltonian}
\end{align}
Here, $\Gamma_{uu}$ ($\Gamma_{ll}$) is the energy bias that need to be applied when both electrons are found in the upper (lower) dot. Note that here, unlike for \eqref{eq:X-_Hamiltonian}, we chose an energy reference, the (1,1)-configuration has a fixed energy of 0. Without external magnetic field, the three triplet states are degenerate. Again, we obtain voltage-dependent transition energies when diagonalizing the Hamiltonians \eqref{eq:X2m_Hamiltonian} and \eqref{eq:e2-_Hamiltonian} and subtracting the resulting spectra. We can again use the model to identify the transitions in the PL-V  spectrum and determine the QDM parameters for the $X^{2-}$. The PL-V overlaid with the transition energies obtained from fitting the PL-V with the model is shown in Fig.\,\ref{fig:SI_X2m_PLV}. Once the parameters are obtained, we can look at the voltage-dependent eigenenergies of the two-electron state, that we get from diagonalizing the Hamiltonian\,\eqref{eq:e2-_Hamiltonian}. These eigenenergies are shown in Fig\,\ref{fig:SI_X2m_PLV}(b). The energy scale is chosen to be zero, if the electron occupy different QDs. One can see that around $V_G = 1\,\mathrm{V}$ the orbital wavefunctions of the electrons hybridize over the two QDs and the well-known singlet-triplet ground states. This regime is also commonly referred to as ``sweet spot'' and it is marked as a blue area in Fig.\,\ref{fig:SI_X2m_PLV}(a) and (b). The parameters dor the doubly charged exciton found from the PL-V are listed in Table\,\ref{tab_X2m}
\begin{table}[ht]
    \caption{Doubly charged exciton parameters determined from voltage-dependent photoluminescence.}
    \label{tab_X2m}
    \begin{tabular}{c|c}
         Parameter & Value \\\hline
         $E_{X^{2-}}$ & 1336.0(1)\,meV\\
         $\Gamma_{X^{2-}}$ & 25.2(1)\,meV\\
         $\Gamma_{ll}$ & 20.5(1)\,meV\\
         $\Gamma_{uu}$ & -9.2(1)\,meV
    \end{tabular}
\end{table}

\section{\label{sec:setup}Experimental setup}
\begin{figure}
    \includegraphics[width=0.7\linewidth]{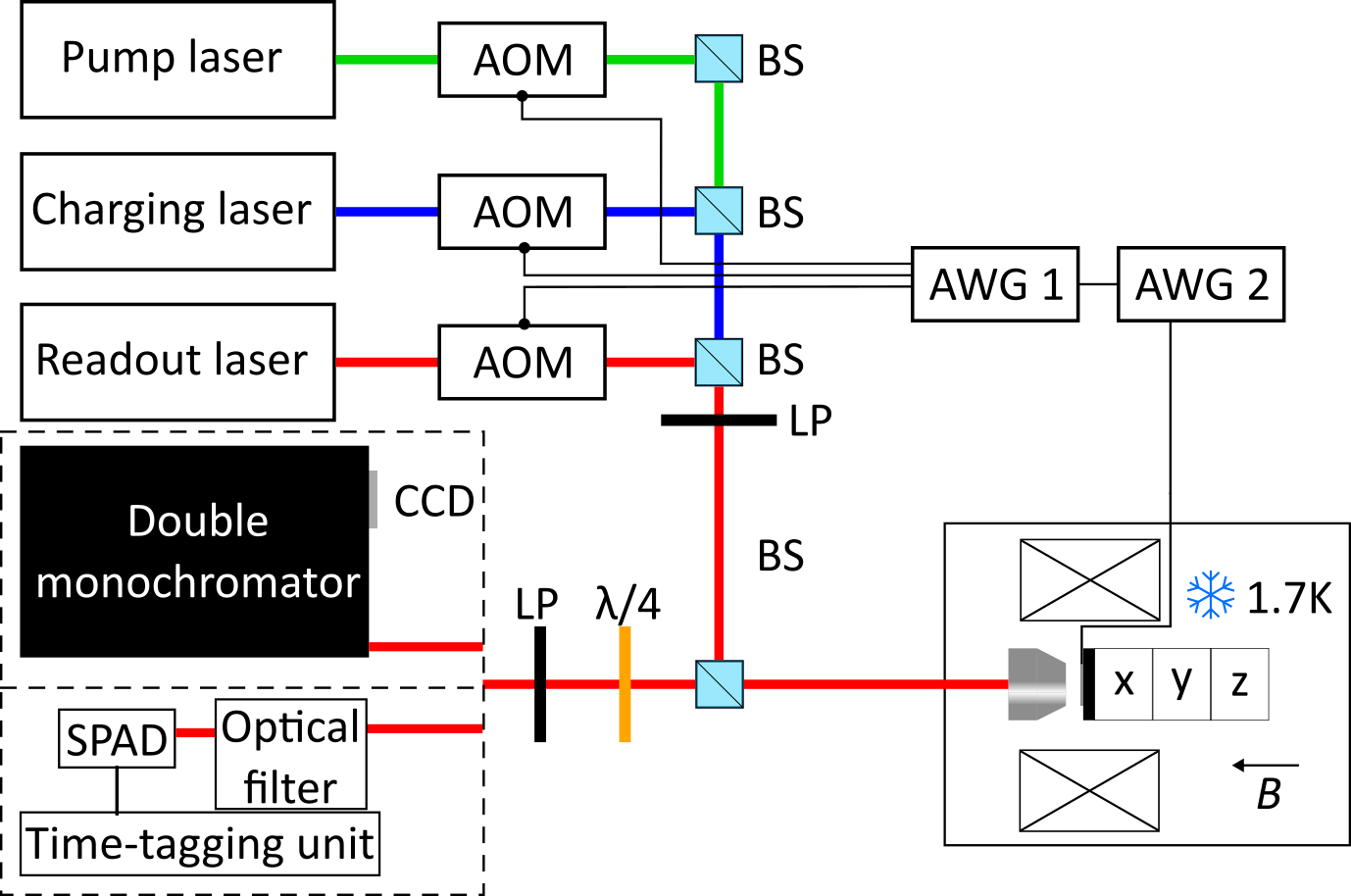}
    \caption{\label{fig:setup}Schematic of the experimental setup. Three continuous wave lasers are passed through acousto-optic modulators (AOMs). The AOMs are gated using an arbitrary waveform generator (AWG 1). The lasers are combined using beam splitters (BS) and guided to the sample that is built into a cryostat (Attocube2100) and held at 1.7\,K. The sample is placed on nanopositioners (Attocube ANP). A second AWG sets the voltage sequence that is applied to the QDM. Via a superconducting solenoid magnet, a magnetic field of up to 9\,T can be applied in the growth direction of the sample (Faraday geometry). Luminescence photons of the dot are collected and either analyzed using a double monochromator and a liquid nitrogen-cooled CCD to obtain spectral data, or, after optical filtering, can be analyzed in the time domain using a single photon avalanche diode (SPAD) and subsequent time tagging unit. Polarization control and suppression of the excitation laser and the emitted photons is facilitated by using linear polarizers (LPs) and a quarter wave plate ($\lambda/4$).}
\end{figure}

The experimental setup used for the charging experiments is illustrated schematically in Fig.\,\ref{fig:setup}. The lasers used for charging, readout and pumping are tunable diode lasers (two Toptica CTL, one Toptica DLPro) are gated using acousto-optic modulators (AA Opto-electronic MT250-IR6). They receive their signal from an arbitrary waveform generator (Tektronix AWG5204) that furthermore triggers the signal output of a second waveform generator (Keysight 33500B) to gate the sample diode, thus synchronizing the lasers and the gate voltage signal. We use a confocal microscope with an 0.81\,NA objective lens to achieve laser excitation and fluorescence collection. To filter out resonant laser light, we use cross-polarized detection of the emitted photons. Not shown in the figure is a fourth laser emitting at 850\,nm used to measure voltage-dependent photoluminescence via excitation of charge carriers in the wetting layer (see section\,\ref{sec:plv}). It is possible to either measure spectral properties of the emitted photons using a double monochromator and liquid nitrogen cooled CCD, or to perform time-resolved measurements by counting (spectrally filtered) photons using a single photon avalanche diode (Excelitas SPCM-AQRH-16-FC) and time-correlating them with the waveform signals using a time tagging unit (Swabian Instruments Time Tagger Ultra).

\section{\label{sec:charge_plateaus}Determining charging parameters}
\begin{figure}[ht]
    \includegraphics[width=0.5\linewidth]{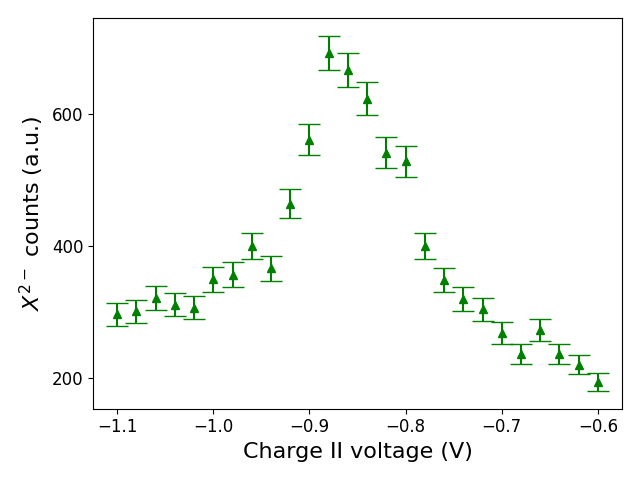}
    \caption{\label{fig:2nd_chargeV}$X^{2-}$ intensity vs. second charge voltage plateau with the charge laser energy fixed at 1361.0\,meV.}
\end{figure}

For charging the QDM with on electron, we used a charge laser energy of 1361.0\,meV. This will charge the QDM with one electron at a gate voltage of -4\,V during the first charging phase. To charge the QDM with a second electron, we use a sequential charging method, where we keep the charge laser energy constant but vary the charge voltage. We found the optimal voltage to charge a second electron to be at $-0.88$\,V.\\
\begin{figure}[ht]
    \includegraphics[width=0.8\linewidth]{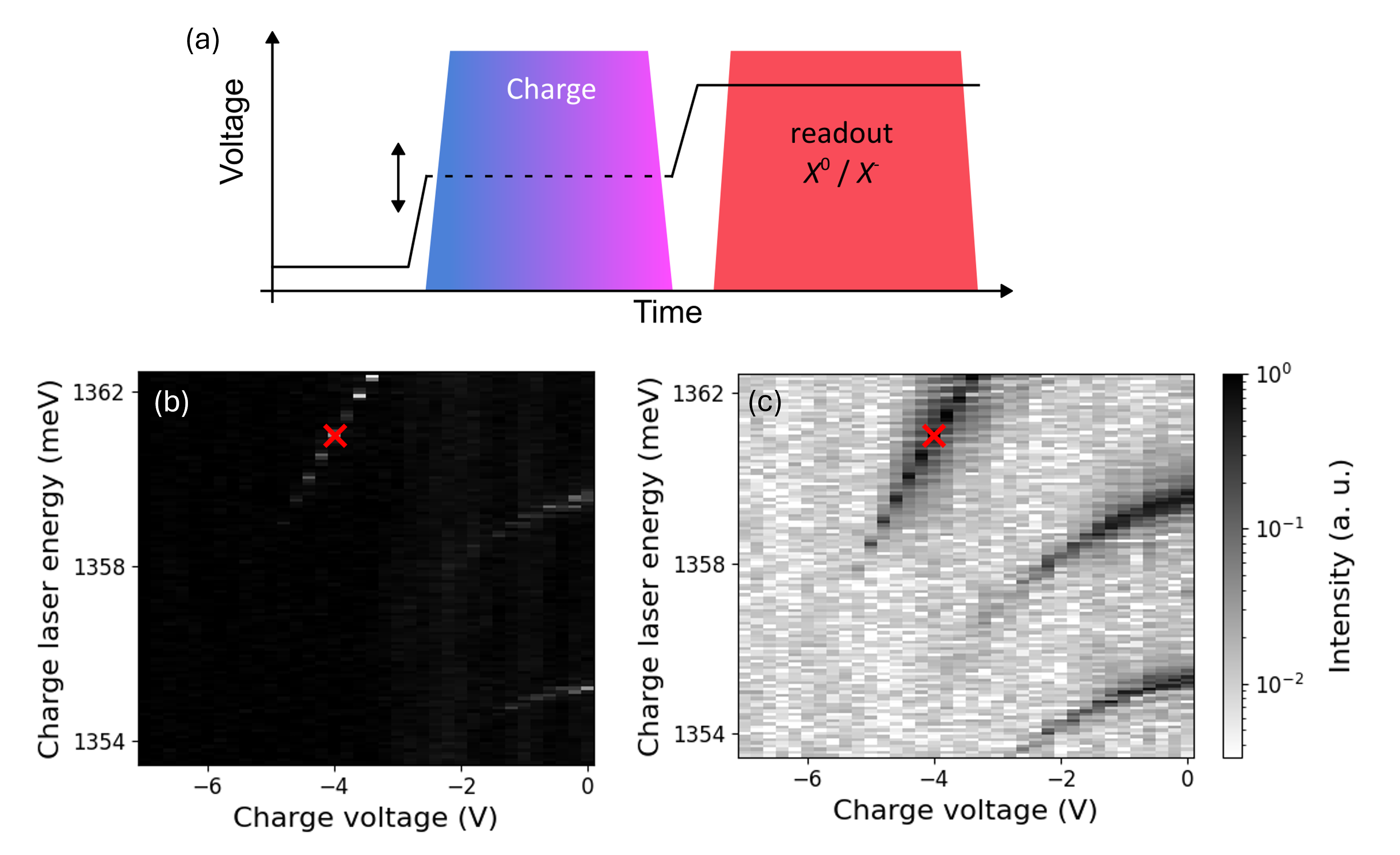}
    \caption{\label{fig:charge_maps}(a) Experimental cycle used to determine the charging parameters for one-electron-charging. (b) $X^0$ intensity map as function of charge voltage and charge laser energy. The cross marks the working point used for subsequent experiments in this manuscript. (c) $X^-$ intensity map as function of charge voltage and charge laser energy.}
\end{figure}
In this section, we explain how we found these parameters. In a first step, we determine the optimal charge laser energy and charge voltage. For this, we employ the measurement cycle shown in Fig.\,\ref{fig:charge_maps}(a), where the readout laser is set resonant with $X^0$. As a readout voltage we typically choose 1\,V since in this regime all three charged complexes under investigation are reasonably close together, yet still well distinguishable (see also Fig.\,\ref{fig:plv}(c)) We then scan the two-dimensional parameter space by changing the charge laser energy (indicated by the color gradient of the laser pulse in Fig.\,\ref{fig:charge_maps}(a)) and the charge laser voltage (indicated by the arrows next to the charge voltage level in Fig.\,\ref{fig:charge_maps}(a)). If no charging takes place, there will be a strong $X^0$ signal visible during the readout. However, if charging takes place, this signal should quench, since the readout laser is not resonant with the $X^-$. In Fig.\,\ref{fig:charge_maps}(b), we observe that for most combinations of charge laser energy and charge voltage, the $X^0$ remains bright. But there are some valleys where the $X^0$ goes dark, which could be indicative of charging taking place.\\
To show that this is actual charging with one electron, we perform the same experiment, but with readout of $X^-$. The results of these measurements are shown in \ref{fig:charge_maps}(c). One can see that the $X^-$ turns bright exactly where the $X^0$ goes dark. This is a clear sign that the observed loss of the $X^0$ signal is due to one-electron charging.\\
As a working point for one-electron charging we chose -4\,V as charging voltage and 1361.0\,meV for the charge laser energy (marked as red crosses in Fig.\,\ref{fig:charge_maps}(b) and (c)). For two-electron charging, we keep the charge laser on at the same energy, but we vary the charge voltage, so that the charge laser is resonant with an $X^-$ transition. This will excite a trion, of which the hole tunnels out and the QDM is left with two electrons. To find the charge charging voltage, we perform a similar experiment as before but now we keep the charge laser energy fixed and only sweep the second charge voltage. Furthermore, we set the readout laser resonant with the $X^{2-}$ transition to probe two-electron-occupation. The result of this measurement is shown in Fig.\,\ref{fig:2nd_chargeV}. We observe a peak in the $X^{2-}$ emission at around $-0.88$\,V. We used this voltage to charge the QDM with a second electron for all experiments involving two electrons.

\section{\label{sec:charge_fidelities}Estimating charging fidelities}
To estimate the charge fidelities, we follow the method in\,\cite{bopp_quantum_2022}. We denote the probability to charge $n$ electrons given that $m$ electrons were intended to be charged (i.e. the $m$ is the number of charge plateaus) as $p(n|m)$. Our starting point is the assumption that the QDM is initialized (during the reset phase) to be perfectly empty, i.e. $p(0|0)=1$. We furthermore restrict our analysis to the lowest three charge states $X^0$, $X^-$ and $X^{2-}$. We denote the number of counts of charge state $a\in\{0, -, 2-\}$ given that $b\in\{0,1,2\}$ charge pulses were applied as $N^a_b$.\\
Now, the probability of charging one electron given that one pulse was applied is $p(1|1)=p(\neg 0|1) - p(2|1)$ where the probability to charge more than one electron given that one charge pulse was applied is
\begin{equation}
    p(\neg 0|1) = \frac{N_0^0-N_1^0}{N_0^0}
\end{equation}
and $p(2|1)\leq \frac{N_1^{2-}}{N_2^{2-}}$. Thus, the one-electron charging fidelity can be written as
\begin{equation}
    p(1|1) \geq \frac{N_0^0-N_1^0}{N_0^0}-\frac{N_1^{2-}}{N_2^{2-}}.
    \label{eq:p11}
\end{equation}
Using the count numbers from the sequential charging experiment presented in\,\ref{sec:charging}, we determine a value of $p(1|1)\geq85.9(\pm0.2)\%$ from Eqn.~\eqref{eq:p11}.
For the two-electron charging fidelity we can use $p(2|2)=1-p(1|2)-p(0|2)$ where $p(0|2)\leq 1-p(\neg 0|1)$ and $p(1|2)\leq \frac{N_2^{2-}}{N_1^-}$. We arrive at the following expression for the two-electron charging fidelity:
\begin{equation}
    p(2|2)\geq \frac{N_0^0-N_1^0}{N_0^0}-\frac{N_2^{-}}{N_1^{-}}.
\end{equation}
We find $p(2|2)\geq89.2(\pm11.1)\%$ for our measured data.

\section{\label{sec:selection_rules}Selection rules of the doubly charged exciton}
\begin{figure}[ht]
    \centering
    \includegraphics[width=0.3\textwidth]{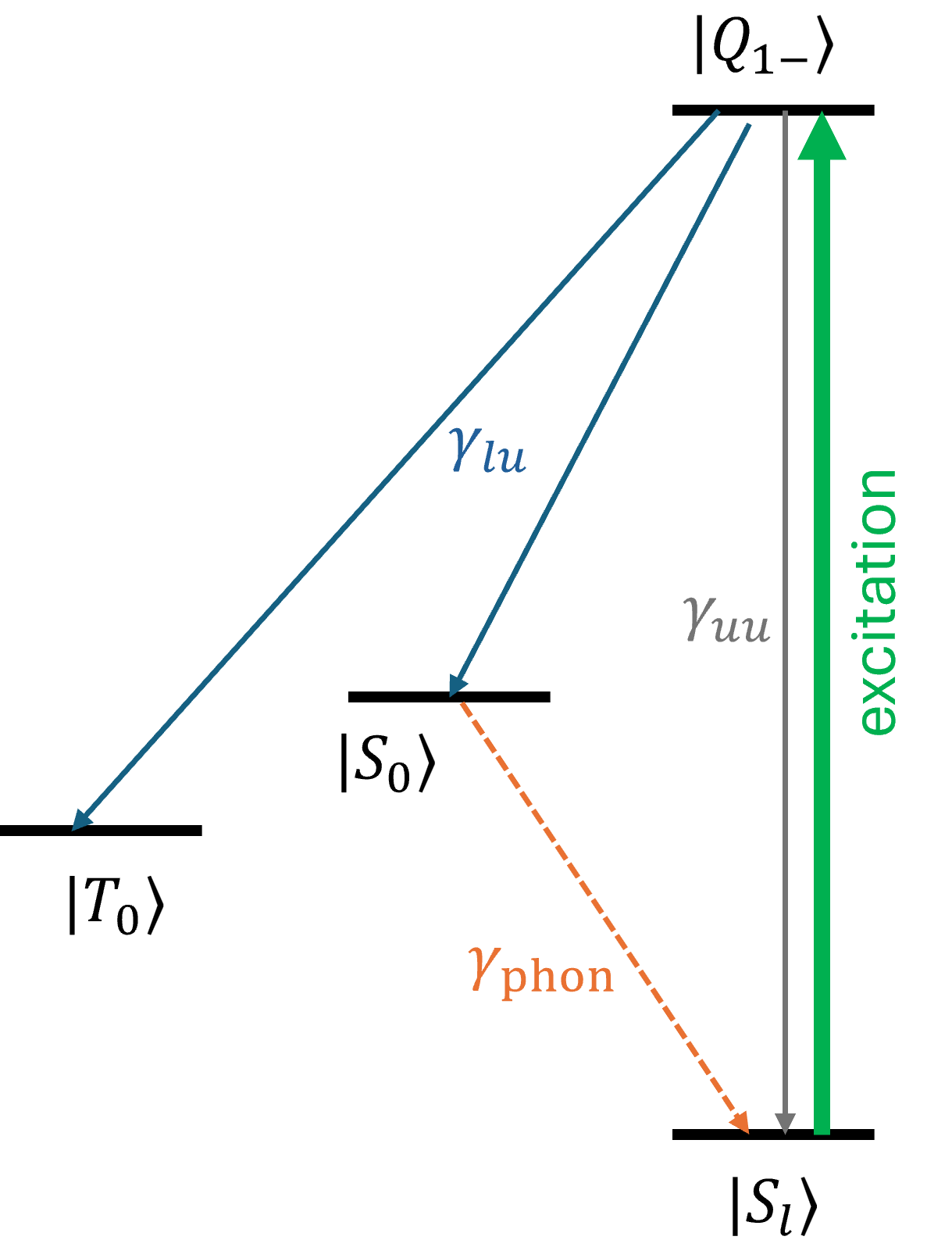}
    \caption{Level scheme for the $2e$ pumping of the $\ket{T_0}$ state.}\label{fig:2e_pumping}
\end{figure}

In this section, we derive the selection rules between the two-electron states and the doubly charged exciton states. While the selection rules in the (1,1)-regime with applied magnetic field in growth direction, as presented in Fig.\,\ref{fig:charge_storage}(d), can be understood intuitively from the change of the net spin in the upper dot\,\cite{delley_deterministic_2017, weiss_coherent_2012}, the selection rules in the (2,0)-regime, where most experiments in section\,\ref{sec:spin_pumping} were performed, are more complicated. Here, the trion state in terms of charge occupation is described by
\begin{equation}
    \begin{pmatrix}
        2 & 1 \\
        0 & 1
    \end{pmatrix}.
\end{equation}
The state consists of an electron pair in the lower dot and an electron-hole-pair in the upper dot. With regards to spin, this charged complex should therefore behave like a neutral exciton confined in the upper dot and therefore the electron-hole-exchange interaction becomes an important factor, that governs the selection rules for both, the direct and indirect transition\,\cite{bayer_fine_2002}.\\
We base our theoretical calculation of the selection rules on the description in second quantization using the single electron states $\ket{\psi_i}$, where $i = \{ \lambda_i, \alpha_i, \sigma_i \}$ is a compound quantum number describing the QD location $\alpha_i = (l, u)$ (for lower and upper QD), the band index $\lambda_i = (e, h)$ (for electrons and holes) and the spin $\sigma_i = (\uparrow, \downarrow)$. The light-matter interaction Hamiltonian, which governs the selection rules, is given by 
\begin{align}
    H_\mathrm{LM} = - \ve{d} \cdot \ve{E}(\ve{r_0}, t)
\end{align}
in the dipole approximation. $\ve{d} = - e \ve{r}$ is the dipole operator (with elementary charge $e$ and position operator $\ve{r}$) and $\ve{E}(\ve{r_0}, t)$ is the time-dependent electric field at emitter position $\ve{r_0}$. A monochromatic electric field of frequency $\omega$ can be decomposed into
\begin{align}
    \ve{E} (\ve{r_0}, t) \equiv \ve{E}^{(+)} \ee^{ - \ii \omega t} +  \ve{E}^{(-)} \ee^{\ii \omega t}
\end{align}
with  $\ve{E}^{(+)}$ ($\ve{E}^{(-)}$) being the positive (negative) frequency part of the electric field, which are the complex conjugates of each other. In the rotating wave approximation and second quantization, the Hamiltonian can be rewritten into
\begin{align}
    H_\mathrm{LM} = - \sum_k E_k^{(+)} \sum_{i,j}\bra{\psi_i} d^k \ket{\psi_j} e_i^\dagger h_j^\dagger + \mathrm{h.c.}.
    \label{eq:hamiltonian_with_me}
\end{align}
$E_k^{(+)} = \ve{E}^{(+)} \cdot \ve{e}_k$ is the projection of the positive-frequency electric field onto the $k$-th polarization basis vector and $d^k = \ve{d} \cdot \ve{e}_k^\ast$, where $e_i$ ($h_j$) is the annihilation operator of an electron (a hole) in state $\ket{\psi_i}$ ($\ket{\psi_j}$) and the sum over $i$ ($j$) only runs over electron (hole) states. \\
To calculate the many-body matrix elements, we need the single-particle dipole matrix elements $\bra{\psi_i} d^k \ket{\psi_j}$. In the envelope function approximation, the wave functions are written as a product of the Bloch factor $u_{\lambda_i, \sigma_i} (\ve{r})$ and a slowly varying envelope function $\varphi_{\alpha_i}(\ve{r})$\,\cite{haug2008quantum,chow2013physics}:
\begin{align}
    \psi_i(\ve{r}) = u_{\lambda_i, \sigma_i} (\ve{r})\varphi_{\alpha_i}(\ve{r}).
\end{align}
Then, the dipole matrix elements can be approximated as
\begin{align} \label{eq:efa_matrix_elements}
    \bra{\psi_i} d^k \ket{\psi_j} \approx \int \dd^3 r \, u_{\lambda_i, \sigma_i}^\ast (\ve{r}) d^k u_{\lambda_j, \sigma_j}(\ve{r}) \int \dd^3 r \, \varphi_{\alpha_i}^\ast(\ve{r}) \varphi_{\alpha_j}(\ve{r}),
\end{align}
meaning that the selection rules are governed by the bulk dipole matrix elements and the envelope functions only influence the amplitude of the transition. The spin-dependence of the Bloch factors is due to spin-orbit coupling. In InGaAs, the spin and orbital angular momentum of heavy holes are aligned. The hole Bloch factors have $p$-type symmetry (angular momentum $l=1$) and we neglect light holes (non-aligned spin and orbital angular momentum), since they are higher in energy. The conduction band electrons have $s$-type symmetry ($l=0$) and therefore don't exhibit direct spin-orbit coupling\,\cite{bayer_fine_2002,gywat2010spins}. For circularly polarized light ($\ve{e}_{\sigma^+} = \frac{1}{\sqrt{2}}(1, \, \ii)^T $, $\ve{e}_{\sigma^-} = \frac{1}{\sqrt{2}}(1, \, -\ii)^T$), the matrix element selection rules can be inferred from symmetry considerations or by evaluating the angular part of the first integral in Eq.~\eqref{eq:efa_matrix_elements}. The only non-vanishing single-particle matrix elements are
\begin{align}
    \bra{\psi_{\mathrm{e}, \alpha, \downarrow}} d^{\sigma^+} \ket{\psi_{\mathrm{h}, \beta, \uparrow}}  = 
    \bra{\psi_{\mathrm{e}, \alpha, \uparrow}} d^{\sigma^-} \ket{\psi_{\mathrm{h}, \beta, \downarrow}} \equiv D_{\alpha, \beta} = D^{(0)}\chi_{\alpha, \beta}.
\end{align}
Their amplitudes are given by the bulk dipole matrix element $D^{(0)}$ and the envelope function overlap $\chi_{\alpha, \beta} =  \int \dd^3 r \, \varphi_{\alpha}^\ast(\ve{r}) \varphi_{\beta}(\ve{r})$. Therefore, direct transitions ($\alpha = \beta$) do have stronger transition amplitudes than indirect ($\alpha \neq \beta$) ones. To calculate the transition amplitudes, we first define all relevant states in terms of creation operators. We consider all two-electron states
\begin{align}
    \ket{S_0} &= \frac{1}{\sqrt{2}} (e_{l, \uparrow}^\dagger e_{u, \downarrow}^\dagger - e_{l, \downarrow}^\dagger e_{u, \uparrow}^\dagger) \ket{\mathrm{vac}}, \nonumber\\
    \ket{T_0} &= \frac{1}{\sqrt{2}} (e_{l, \uparrow}^\dagger e_{u, \downarrow}^\dagger + e_{l, \downarrow}^\dagger e_{u, \uparrow}^\dagger) \ket{\mathrm{vac}}, \nonumber\\
    \ket{S_u} &= e_{u, \uparrow}^\dagger e_{u, \downarrow}^\dagger \ket{\mathrm{vac}}, \nonumber\\
    \ket{S_l} &= e_{l, \uparrow}^\dagger e_{l, \downarrow}^\dagger \ket{\mathrm{vac}}, \nonumber\\
    \ket{T_-} &= e_{l, \downarrow}^\dagger e_{u, \downarrow}^\dagger \ket{\mathrm{vac}}, \nonumber\\
    \ket{T_+} &= e_{l, \uparrow}^\dagger e_{u, \uparrow}^\dagger \ket{\mathrm{vac}}.
    \label{eq:2e_states}
\end{align}
For the doubly charged exciton states, we will only focus on the aforementioned $\begin{pmatrix}
        2 & 1 \\
        0 & 1
    \end{pmatrix}$-states, for which we will introduce the shorthand notation $\ket{Q}$. With electron-hole-exchange-interaction of the exciton in the upper dot included, the eigenstates at zero magnetic field can be expressed in terms of creation operators as
\begin{align}
    \ket{Q_{1\pm}} &= \frac{1}{\sqrt{2}} e_{l, \uparrow}^\dagger e_{l, \downarrow}^\dagger (e_{u, \downarrow}^\dagger h_{u, \uparrow}^\dagger \pm e_{u, \uparrow}^\dagger h_{u, \downarrow}^\dagger) \ket{\mathrm{vac}}, \nonumber\\
    \ket{Q_{2\pm}} &= \frac{1}{\sqrt{2}}e_{l, \uparrow}^\dagger e_{l, \downarrow}^\dagger (e_{u, \uparrow}^\dagger h_{u, \uparrow}^\dagger \pm e_{u, \downarrow}^\dagger h_{u, \downarrow}^\dagger) \ket{\mathrm{vac}}.
    \label{eq:trion_states}
\end{align}
The eigenstates are two well-known bright states with total spin projection $M=1\pm$ and two dark states with total spin projection $M=2\pm$.\\
Now, we consider the basis given in Eq.~\eqref{eq:2e_states} and the last four states in Eq.~\eqref{eq:trion_states} in that order. The dipole Hamiltonian in this basis is given by
\begin{align}
\begin{aligned}
H_{\mathrm{LM}}
&= E_{\sigma^+}^{(+)}\Bigg[
\frac{d_{lu}}{2}\Big(
\ket{Q_{1+}}\bra{S_0}
+ \ket{Q_{1+}}\bra{T_0}
+ \ket{Q_{1-}}\bra{S_0}
+ \ket{Q_{1-}}\bra{T_0}
\Big) \\
&\quad
+ \frac{d_{uu}}{\sqrt{2}}\Big(
\ket{Q_{1+}}\bra{S_l}
+ \ket{Q_{1-}}\bra{S_l}
\Big)
+ \frac{d_{lu}}{\sqrt{2}}\Big(
\ket{Q_{2+}}\bra{T_+}
+ \ket{Q_{2-}}\bra{T_+}
\Big)
\Bigg]\\
&\quad+ E_{\sigma^-}^{(+)}\Bigg[
\frac{d_{lu}}{2}\Big(
-\ket{Q_{1+}}\bra{S_0}
+ \ket{Q_{1+}}\bra{T_0}
+ \ket{Q_{1-}}\bra{S_0}
- \ket{Q_{1-}}\bra{T_0}
\Big) \\
&\quad
+ \frac{d_{uu}}{\sqrt{2}}\Big(
\ket{Q_{1+}}\bra{S_l}
- \ket{Q_{1-}}\bra{S_l}
\Big)
+ \frac{d_{lu}}{\sqrt{2}}\Big(
\ket{Q_{2+}}\bra{T_+}
- \ket{Q_{2-}}\bra{T_+}
\Big)
\Bigg] + \mathrm{h. c.}
\end{aligned}
\end{align}
which can be transformed into the linear representation
\begin{align}
\begin{aligned}
H_{\mathrm{LM}}
&= E_{\mathrm{h}}^{(+)}\Bigg[
\frac{d_{lu}}{\sqrt{2}}\Big(
\ket{Q_{1+}}\bra{T_0}
+ \ket{Q_{1-}}\bra{S_0}\Big) + d_{uu}\ket{Q_{1+}}\bra{S_l} \\
&\quad
+ \frac{d_{lu}}{2}\Big(
\ket{Q_{2+}}\bra{T_-} + \ket{Q_{2+}}\bra{T_+}-\ket{Q_{2-}}\bra{T_-} + \ket{Q_{2-}}\bra{T_+}
\Big)
\Bigg]\\
&\quad- \mathrm{i}E_{\mathrm{v}}^{(+)}\Bigg[
\frac{d_{lu}}{\sqrt{2}}\Big(
\ket{Q_{1+}}\bra{S_0}
+ \ket{Q_{1-}}\bra{T_0}\Big) + d_{uu}\ket{Q_{1-}}\bra{S_l} \\
&\quad
+ \frac{d_{lu}}{2}\Big(
-\ket{Q_{2+}}\bra{T_-} + \ket{Q_{2+}}\bra{T_+}+\ket{Q_{2-}}\bra{T_-} + \ket{Q_{2-}}\bra{T_+}
\Big)
\Bigg] + \mathrm{h. c.}
\end{aligned}
\end{align}
by using 
\begin{align}
    E_\mathrm{h} &= \frac{1}{\sqrt{2}} (E_{\sigma^+} + E_{\sigma^-}), \nonumber \\
    E_\mathrm{v} &= \frac{\ii}{\sqrt{2}} (E_{\sigma^+} - E_{\sigma^-}).
\end{align}
\begin{figure}[ht]
    \centering
    \includegraphics[width=0.35\textwidth]{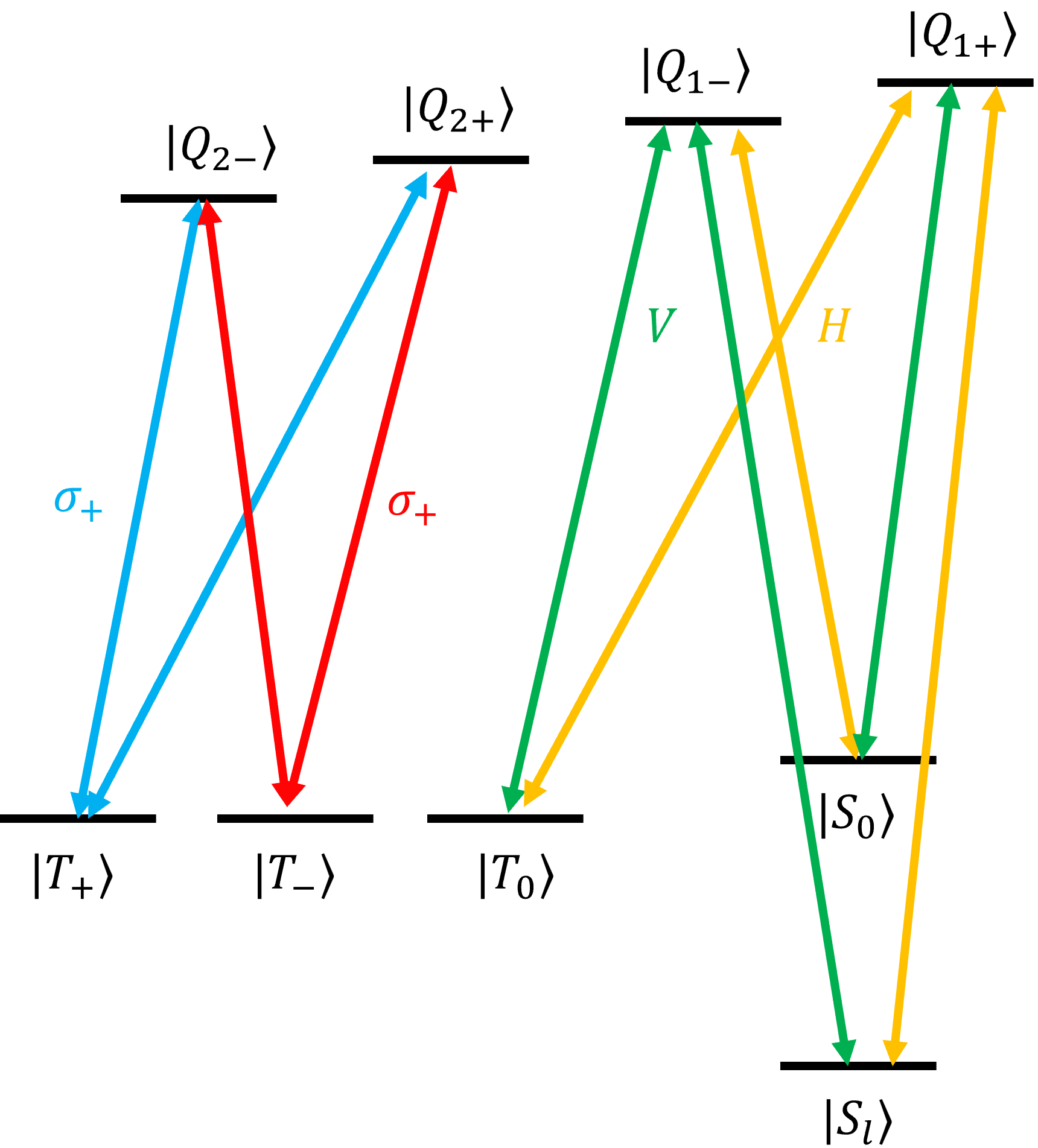}
    \caption{Level scheme and optical transitions between $\ket{Q}$ states and the two-electron states. Linearly polarized transitions are displayed in blue (vertical) and red (horizontal), while circularly polarized transitions are displayed in orange ($\sigma^+$) and green ($\sigma^-$).}\label{fig:transitions}
\end{figure}
The selection rules that can be inferred from this Hamiltonian are displayed as a level scheme in Fig.~\ref{fig:transitions}. The states $\ket{Q_{1\pm}}$ couple to $\ket{S_0}$, $\ket{T_0}$ and $\ket{S_l}$ via linearly polarized photons, while $\ket{Q_{2\pm}}$ each couple to the remaining triplet states via circularly polarized photons. This leads to a formation of two subspaces, that do not couple optically.
\\
In our experiments we optically drive the transition $\ket{S_B}\leftrightarrow\ket{Q_{1-}}$ with a narrow band cw-laser, such that the other trion states can be safely neglected.
\\
The scheme of $\ket{T_0}$-pumping is shown Fig.~\ref{fig:2e_pumping}. With the state initialized into the $\ket{S_l}$ state, it is pumped into the $\ket{Q_{1-}}$ state via linearly polarized light. From there, $\ket{Q_{1-}}$ can decay back into $\ket{S_l}$, which is however offset by the continuous pumping. A portion of the population decays into both $\ket{S_0}$ and $\ket{T_0}$ with roughly equal rate $\gamma_\mathrm{lu}$. The delocalized singlet $\ket{S_0}$ quickly decays back into $\ket{S_l}$ due to phonon-assisted tunneling ($\gamma_\mathrm{phon}$) on the order of picoseconds, similar to fast interdot-tunneling of a single electron, and is the re-excited by the pump laser. The population in $\ket{T_0}$ stays there for a longer time since phonon-assisted tunneling preserves spin and therefore only allows transitions between singlet and triplet states due to small contributions stemming from Dresselhaus spin-orbit coupling and other spin-flip processes, which are however much slower\,\cite{gawelczyk2021tunnelingrelated,gawarecki_phonon-assisted_2021}.\\
It should be noted that this simple analytic derivation does not take into account various aspects of the actual quantum dot molecule, such lateral misalignment of the dots, different lateral asymmetry of the two dots, heavy-light-hole mixing and different strain environments in each dot, all of which are expected to slightly change the idealized selection rules shown in Fig.\,\ref{fig:transitions}\,\cite{weiss_coherent_2012, farfurnik_single-shot_2021}.
\section{\label{sec:theory_kp}The phonon-driven quantum kinetics of the singlet-triplet configurations}

In this section, we consider the triplet $T_0$ to the singlet $S$ phonon-assisted transitions. The modeling is based on the eight-band $\bm{k}\cdot \bm{p}$ theory and involves structural calculations in the real space.

\subsection{The structural calculations}
\begin{figure}[tb]
    \centering
    \includegraphics[width=0.5\linewidth]{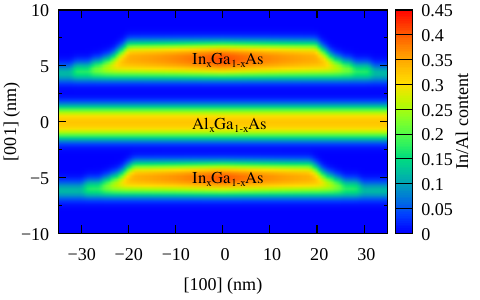}
    \caption{The indium distribution (In$_{x}$Ga$_{1-x}$As) in the QDs and Al content (Al$_{x}$Ga$_{1-x}$As) in the barrier.}
    \label{fig:qd_geometry_theory}
\end{figure}
The shapes of the quantum dots are modeled using truncated Gaussian profiles~\cite{Lienhart2025}. The top surface of each dot is described by \[ S(x,y) = \min\!\left( w\,\exp\!\left[-\frac{x^{2}+y^{2}}{2d^{2}}\right],\, h \right), \] where \(h\) denotes the dot height, \(d\) controls the lateral size, and \(w\) sets the Gaussian steepness. For the lower and upper dots (indexed by \(l\) and \(u\)), the parameters are \(h_l = 3.5a\), \(d_l = 48a\), \(w_l = 25a\), and \(h_u = 4a\), \(d_u = 52a\), \(w_u = 25a\), with \(a\) being the lattice constant. Both dots are placed on wetting layers of thickness \(a\). The separation between the dots (base to base) is \(17.5a\), and the Al\(_{0.33}\)Ga\(_{0.67}\)As barrier between them has a thickness of \(4.5a\).

To describe the individual QD morphology, we employ the trumpet-shaped composition profile~\cite{Jovanov2012,Migliorato2002}, in which the local indium concentration inside the quantum dots is described by
\begin{align*}
    C(\bm{r}) = C_l + \qty(C_u - C_l) \exp{-\frac{\sqrt{x^2+y^2} \exp{-z/z_0}}{r_0}}.
\end{align*}
Here, $C_u$ and $C_l$ denote the maximum and minimum indium contents in the QD, respectively, while \(r_0\) and \(z_0\) determine the spatial extent of the profile. For both quantum dots, we set \(C_u = 0.42\) (In\(_{0.42}\)Ga\(_{0.58}\)As), \(C_l = 0.3\) (In\(_{0.3}\)Ga\(_{0.7}\)As), and \(r_0 = 20a\). The vertical decay parameters were chosen as \(z_{0\mathrm{l}} = 3.5a\) for the lower dot and \(z_{0\mathrm{u}} = 4.0a\) for the upper one. The dots are separated by the distance $D=17.5a$. 
The cross-section of the resulting material distribution is shown in Fig.~\ref{fig:qd_geometry_theory}.

The strain caused by the lattice mismatch between the involved materials is taken into account within the continuous elasticity approach~\cite{Pryor1998b} with the parameters from Vurgaftman et al.~\cite{Vurgaftman2001}. As in the zinc-blende structure shear strain comes with piezoelectric field, we accounted for this effect with polarization up the second order in the strain tensor components~\cite{Bester2006b}. Here, we took the parameters from Caro et al.~\cite{Caro2015}.

\subsection{The energy levels}
\label{sec:kp}
The wave functions for the electron single-particle states are found within the eight-band $\bm{k}\cdot \bm{p}$ model in the envelope function approximation, as described in Refs.~\cite{Gawarecki2018a, Lienhart2025}. To account for the gate electric field effect, the real-space Hamiltonian is supplemented by the diagonal term
\begin{equation*}
    V_{F} = e F (z-z_\mathrm{c}),
\end{equation*}
where $z_\mathrm{c}$ is between QDs.
For $F \leq 17$~kV/cm, we calculate 12 states: six energetically lowest states (the $s$ and $p$-type envelope) in the lower QD and the corresponding six states of the upper QD. For larger fields, we take the $6$ and $2$ lowest states in the lower and in the upper QD, respectively; which is motivated by better numerical stability. Therefore, the $12$ (or $8$) states form a basis for the configuration interaction procedure, which is the next stage of our modeling.

The two-electron states are found from the diagonalization of the Coulomb Hamiltonian using the configuration interaction method. The calculated energy levels are shown in Fig.~\ref{fig:spin_pumping}(d). The simulations were carried out at small magnetic field $B=0.1$~T in the Faraday orientation. As can be seen, the obtained singlet-triplet energy splitting at the sweet spot (here around $F\approx 10$~kV/cm) of $870\,\mu$eV is reasonably close to the experimental value ($718\,\mu$eV). The character of states in the vicinity of the sweet spot was discussed in the main part of the text. Here, we focus on the configurations at higher electric field (e.g. $F \approx 22$~kV/cm), which favors the localization in the lower dot. 
In this regime, the lowest state is the spin singlet localized in the lower QD $\ket{S_l}$. The next two branches are the triplet states $\ket{T'_{-/0/+}}$ and $\ket{T''_{-/0/+}}$, build on the $s$ and $p$ shell states in the lower QD. The next states are the $s$-shell triplet $\ket{T_{-/0/+}}$ and the singlet $\ket{S_{0}}$. The other upper branches are the singlet configurations involving $s$- and $p$-shell in the lower QD, but they will be not taken into account in this analysis. To describe the states of interest, we introduce an extended notation, accounting for the orbital type
\begingroup
\allowdisplaybreaks
\begin{align}
    \ket{S_l} &= e_{l(s), \uparrow}^\dagger e_{l(s), \downarrow}^\dagger \ket{\mathrm{vac}}, \nonumber\\
    \ket{T_-} &= e_{l(s), \downarrow}^\dagger e_{u(s), \downarrow}^\dagger \ket{\mathrm{vac}} \nonumber\\
    \ket{T_0} &= \frac{1}{\sqrt{2}} (e_{l(s), \uparrow}^\dagger e_{u(s), \downarrow}^\dagger + e_{l(s), \downarrow}^\dagger e_{u(s), \uparrow}^\dagger) \ket{\mathrm{vac}}, \nonumber\\
    \ket{T_+} &= e_{l(s), \uparrow}^\dagger e_{u(s), \uparrow}^\dagger \ket{\mathrm{vac}}, \nonumber\\
    \ket{T'_-} &= e_{l(s), \downarrow}^\dagger e_{l(p_1), \downarrow}^\dagger \ket{\mathrm{vac}}, \nonumber\\
    \ket{T'_0} &= \frac{1}{\sqrt{2}} (e_{l(s), \uparrow}^\dagger e_{l(p_1), \downarrow}^\dagger + e_{l(s), \downarrow}^\dagger e_{l(p_1), \uparrow}^\dagger) \ket{\mathrm{vac}}, \nonumber\\
    \ket{T'_+} &= e_{l(s), \uparrow}^\dagger e_{l(p_1), \uparrow}^\dagger \ket{\mathrm{vac}}, \nonumber\\
    \ket{T''_-} &= e_{l(s), \downarrow}^\dagger e_{l(p_2), \downarrow}^\dagger \ket{\mathrm{vac}}, \nonumber\\
    \ket{T''_0} &= \frac{1}{\sqrt{2}} (e_{l(s), \uparrow}^\dagger e_{l(p_2), \downarrow}^\dagger + e_{l(s), \downarrow}^\dagger e_{l(p_2), \uparrow}^\dagger) \ket{\mathrm{vac}}, \nonumber\\
    \ket{T''_+} &= e_{l(s), \uparrow}^\dagger e_{l(p_2), \uparrow}^\dagger \ket{\mathrm{vac}}, \nonumber\\
    \ket{S_0} &= \frac{1}{\sqrt{2}} (e_{l(s), \uparrow}^\dagger e_{u(s), \downarrow}^\dagger - e_{l(s), \downarrow}^\dagger e_{u(s), \uparrow}^\dagger) \ket{\mathrm{vac}}, 
    \label{eq:2e_states_extended}
\end{align}
\endgroup
where $e_{l(\alpha), \uparrow/\downarrow}^\dagger$  ($e_{u(\alpha), \uparrow/\downarrow}^\dagger$) is the creation operator of the $\alpha$-th electron orbital state with $\uparrow/\downarrow$ spin in the lower (upper) dot.

\subsection{Phonon-assisted transitions}

We calculated phonon-assisted relaxation rates between the two-electron energy levels within Fermi's golden rule
\begin{align*}
    \gamma_\mathrm{phon}(i \rightarrow j) = \frac{2\pi}{\hbar} \sum_{\bm{k},\lambda} \abs{G_{ij,\lambda}(\bm{k})}^2 \qty[ \delta( \Delta E_{ij} - \hbar c_\lambda k) + \delta( \Delta E_{ij} + \hbar  c_\lambda k) ],
\end{align*}
where $\Delta E_{ij} = E_i - E_j$ is the energy difference between the energy levels, $G_{ij,\lambda}(\bm{k})$ is the two-electron form-factor, as described in Ref.~\cite{gawarecki_phonon-assisted_2021}; $\lambda$ is the phonon branch (two transversal and a longitudinal one), and $c_\lambda$ is the speed of sound. Here, we take into account the coupling between the electrons and phonons via the deformation potential and the piezoelectric field~\cite{Yu2005,Grodecka2005a,Krzykowski2020}. The material parameters used in calculations are listed in the Appendix of Ref.~\cite{Lienhart2025}. Here, we consider the zero-temperature case, i.e. the transitions from the energetically upper to the lower states (phonon emission). We also note, that we take into account only the spin-flip transitions due to the spin-orbit interaction, neglecting the hyperfine-coupling-related effects. While hyperfine- and spin-orbit–induced spin relaxation originate from different couplings, both mechanisms interact with the same phonon bath and thus share the same phonon spectral density. Consequently, the present model provides an insight into phonon-related transitions, even though the absolute relaxation rates for the two mechanisms may differ substantially.

The calculated direct phonon-assisted transition rate between the $T_0$ and the lowest singlet state $S$ is shown in Fig.~\ref{fig:spin_pumping}(e). The electric field dependence and is how it compares to the experimentally measured ones shown in Fig.\,\ref{fig:spin_pumping}(c) is discussed in section\,\ref{sec:spin_pumping} of the main part of the text.

\end{document}